\documentclass[english,aps,prl,twocolumn,amsmath,amssymb,showpacs,superscriptaddress,notitlepage]{revtex4-2}

\usepackage[T1]{fontenc}
\usepackage{color}
\definecolor{page_backgroundcolor}{rgb}{1, 1, 1}
\pagecolor{page_backgroundcolor}
\usepackage{amsmath}
\usepackage{graphicx}
\usepackage{microtype}
\usepackage{bm}
\usepackage{makecell}
\usepackage[normalem]{ulem}
\usepackage{enumitem}
\usepackage[marginal]{footmisc}
\usepackage{braket}
\usepackage{pifont}
\usepackage{float}

\makeatletter
\usepackage{amssymb}
\usepackage{amsmath}
\usepackage{graphicx}
\usepackage[colorlinks=true,linkcolor=blue,anchorcolor=red,citecolor=blue,urlcolor=blue]{hyperref}
\usepackage[caption=false]{subfig}
\usepackage{lipsum}

\makeatother

\usepackage{babel}

\begin{document}
\global\long\def\figurename{Fig.}%

\title{Spin-Polarized Majorana Modes from Andreev Confinement in an Altermagnetic Strip}

\author{Yang-Yang Li}
\email{yang-yang.li@stonybrook.edu}
\affiliation{Department of Modern Physics, University of Science and Technology of China, Hefei, Anhui 230026, China}
\affiliation{C. N. Yang Institute for Theoretical Physics and Department of Physics and Astronomy, State University of New York at
Stony Brook, Stony Brook, NY 11794-3840, USA}

\author{Song-Bo Zhang}
\email{songbozhang@ustc.edu.cn}
\affiliation{Hefei National Laboratory, Hefei 230088, China}
\affiliation{School of Emerging Technology,
University of Science and Technology of China, Hefei, Anhui 230026, China}

\begin{abstract}
We exploit $d$-wave altermagnetic spin splitting to realize spin-polarized Majorana zero modes through Andreev confinement on a topological-insulator surface with proximity-induced superconductivity. Beneath a finite-width altermagnetic strip, the $d$-wave exchange field shifts the transverse momenta of particle- and hole-like states in opposite directions, allowing subgap quasiparticles to propagate while the surrounding surface remains fully gapped. Repeated Andreev reflection at the strip boundaries confines these quasiparticles, giving rise to quasi-one-dimensional Andreev bands. We derive their quantization condition and show that varying the altermagnetic strength or the orientation of the $d$-wave spin-splitting pattern drives successive transitions between trivial and topological superconducting phases, with the latter hosting one Majorana mode at each strip end. The two modes have opposite spin polarizations, and their localization length depends on the distance to the phase boundaries. The Majorana end states also exhibit distinct transverse spin-density profiles across successive topological regions, evolving from a single peak to multiple peaks.  %Our results establish altermagnetic order as a tunable platform for controlling the topology, spin, and spatial structure of Andreev-confined Majorana modes.
\end{abstract}

\maketitle

\section{Introduction}
\label{sec:intro}

Majorana zero modes (MZMs) are self-conjugate zero-energy Bogoliubov quasiparticles that obey non-Abelian braiding statistics, making them promising building blocks for topological quantum information processing~\cite{ReadGreen2000,Kitaev2001,Ivanov2001,Nayak2008,Alicea2012,Beenakker2013,DasSarma2015,Flensberg2021}. A widely studied strategy for realizing MZMs combines conventional superconductivity, spin-momentum locking, and time-reversal symmetry breaking~\cite{FuKane2008,Lutchyn2010,Oreg2010}. The helical surface states of topological insulators (TIs) provide a natural platform with intrinsic spin-momentum locking, where proximity-induced $s$-wave superconductivity can realize effective spinless $p_x+ip_y$ pairing~\cite{FuKane2008,PotterLee2011,Xu2014,Finck2014,Tikhonov2016,Chen2018}. %Similar ingredients underlie semiconductor-superconductor nanowires, where the interplay of spin-orbit coupling, a Zeeman field and proximity-induced pairing can drive the system into a one-dimensional (1D) topological superconducting phase~\cite{Lutchyn2010,Oreg2010}.
Engineering topological channels on these surfaces offers a route to controlling the location and properties of Majorana end modes.

In particular, Andreev confinement allows such channels to form on a continuous superconducting surface. Papaj and Fu showed that a localized in-plane Zeeman field can create propagating subgap states beneath a narrow strip, while the surrounding superconducting gap confines their transverse motion through Andreev reflection~\cite{PapajFu2021}. Their proposal exploits a gapless state with a segmented Fermi surface that develops in a proximitized spin-helical system under a sufficiently strong in-plane Zeeman field~\cite{YuanFu2018}. Andreev reflections quantize the transverse motion and produce quasi-one-dimensional (1D) Andreev bands, which can enter a topological superconducting phase and host MZMs at the two ends of the strip.

This raises the question of whether Andreev confinement can also be achieved with zero net magnetization. Altermagnetism has recently emerged as a distinct form of collinear compensated magnetic order \cite{Smejkal2022,SmejkalPerspective2022,McClartyRau2024,SongReview2025,ZhuHuoFengZhang2025}. As in a compensated antiferromagnet, oppositely oriented magnetic moments cancel so that the net magnetization vanishes. However, there is no inversion or translation symmetry connecting opposite moments. The electronic bands thus are strongly spin split, a feature more commonly associated with ferromagnets. In altermagnets, this spin splitting varies across momentum space and reverses between symmetry-related regions of the Brillouin zone, as required by the underlying crystal symmetry \cite{Wu2007,Ahn2019,Naka2019,Hayami2019,Smejkal2020,YuanZunger2020,GonzalezHernandez2021,ma2021multifunctional,Shao2021,SmejkalHellenes2022,LiuSpinGroup2022}. Momentum-resolved spin splitting has now been observed in several candidate materials \cite{Lee2024,Krempasky2024,Osumi2024,Reimers2024,Ding2024,Zeng2024CrSb,YangCrSb2025,Jiang2025NatPhys,ZhangF25NP,Regmi2025NatCommun,DeVita2026AdvMater}, and layered insulating $d$-wave altermagnetism has recently been established in La$_2$O$_3$Mn$_2$Se$_2$ \cite{WeiMaterial2025,Asai2026}. The combination of zero net magnetization and large momentum-dependent spin splitting makes altermagnets attractive for magnetic proximity structures. Recent theory has predicted that proximity to an altermagnet can induce momentum-alternating spin splitting in an adjacent nonmagnetic layer~\cite{ZhuProximity2026}. Signatures of this proximity effect have also been reported in experiments on PtTe$_2$/CrSb heterostructures~\cite{Li2026AltermagneticOrder}.

The momentum dependence of altermagnetic spin splitting can have profound consequences in superconducting systems. It can lead to orientation-dependent Andreev reflection \cite{Sun2023,Papaj2023,Zhao2025Orientation,LuSubgap2026}, shift Andreev bound-state spectra \cite{BeenakkerVakhtel2023}, modify Josephson transport \cite{ZhangHuNeupert2024,Ouassou2023,Lu2024,Fukaya2025,SunZhangLiTrauzettel2025,LiHouEtAl2026}, and generate finite-momentum pairing and Bogoliubov Fermi surfaces \cite{ZhangHuNeupert2024,ChakrabortyBlackSchaffer2024,Hong2025,Wei2024Gapless}. Altermagnetism has also been proposed as a route to Majorana end, edge, and corner states in a variety of geometries \cite{Zhu2023,LiLiu2023,Ghorashi2024,Chatterjee2025,Hadjipaschalis2025,Yi2026,Alam2026,Yang2026}. More recently, several works have explored Majorana states in altermagnetic superconducting heterostructures. MZMs have been proposed in systems with zero net magnetization and can be tuned by the orientation of the altermagnetic order \cite{Ghorashi2024,Hadjipaschalis2025,Yang2026}. In 3D altermagnetic TIs, altermagnetism-induced Bogoliubov Fermi surfaces can give rise to topological superconductivity and MZMs~\cite{FuLiTrauzettel2026}. Altermagnetic momentum shifts have also been shown to modify Andreev spectra and propagation phases in Josephson junctions \cite{BeenakkerVakhtel2023,Dong2026PhaseControlled}. In addition, double-peaked Majorana profiles have recently been reported in planar altermagnet-superconductor structures, where they were associated with interface localization induced by anisotropic hopping \cite{SharmaMohanta2026}. These results motivate examining how transverse confinement and strip-end boundary conditions shape the spatial profiles of Majorana end states.

Here we consider a different geometry in which a finite-width altermagnetic strip forms an Andreev-confined channel on an otherwise superconducting TI surface. We focus on a helical surface Dirac cone centered at the non-$\bar{\Gamma}$ time-reversal-invariant momentum $\bar Y$. Starting from the transverse scattering problem, we derive the Andreev quantization condition and show how the altermagnetic strength and orientation control successive gap closings and topological transitions. We then project the Hamiltonian onto the subspace spanned by the lowest particle-hole-related pair of confined states to obtain an effective 1D Dirac model. From this model, we derive the asymptotic Majorana wave functions inside the strip and determine their localization length and spin structure. The two end modes have opposite spin polarizations, and their localization length diverges as the system approaches a topological transition. To examine the transverse structure beyond this lowest-pair description, we consider wave-function matching at the strip ends, where different transverse modes combine coherently. Their interference provides a physical interpretation of the single- and multi-peaked electronic spin-density profiles observed numerically in successive topological regions. Our results thus connect the topological spectrum with the spatial and spin structure of Andreev-confined Majorana states in an altermagnetic proximity structure.

\begin{figure}
\includegraphics[width=0.4\textwidth]{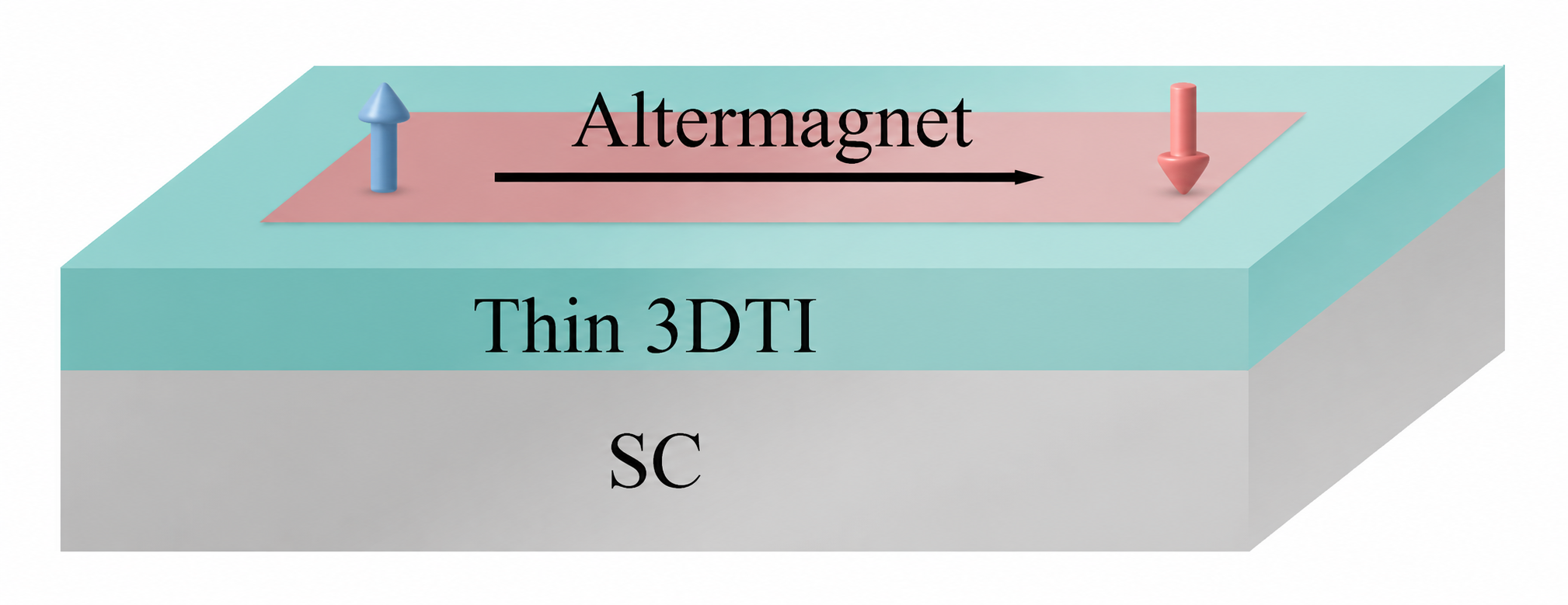}
\caption{Sketch of the setup for realizing MZMs. A TI thin film is proximity-coupled to a superconductor, while an altermagnetic strip on its exposed surface induces a momentum-dependent exchange field. {The black arrow indicates the N\'eel vector of the altermagnetic order.} Andreev confinement produces an MZM at each end of the strip, with opposite electronic spin polarizations indicated by the blue and red arrows. The transverse spin-density profiles evolve from a single peak to multiple peaks across successive topological regions.}
\label{fig:setup}
\end{figure}

\section{Model and Low-Energy States}
\label{sec:model}

We consider a setup shown in Fig.~\ref{fig:setup}, in which an altermagnetic strip of width $W$ is deposited on the superconducting surface of a TI. The strip occupies $|y|<W/2$ and is initially taken to be infinite along $x$, so that $k_x$ remains a good quantum number. We first describe the locally uniform regions inside and outside the strip, and consider a finite strip length when studying the Majorana end states.

\begin{figure}
\includegraphics[width=0.45\textwidth]{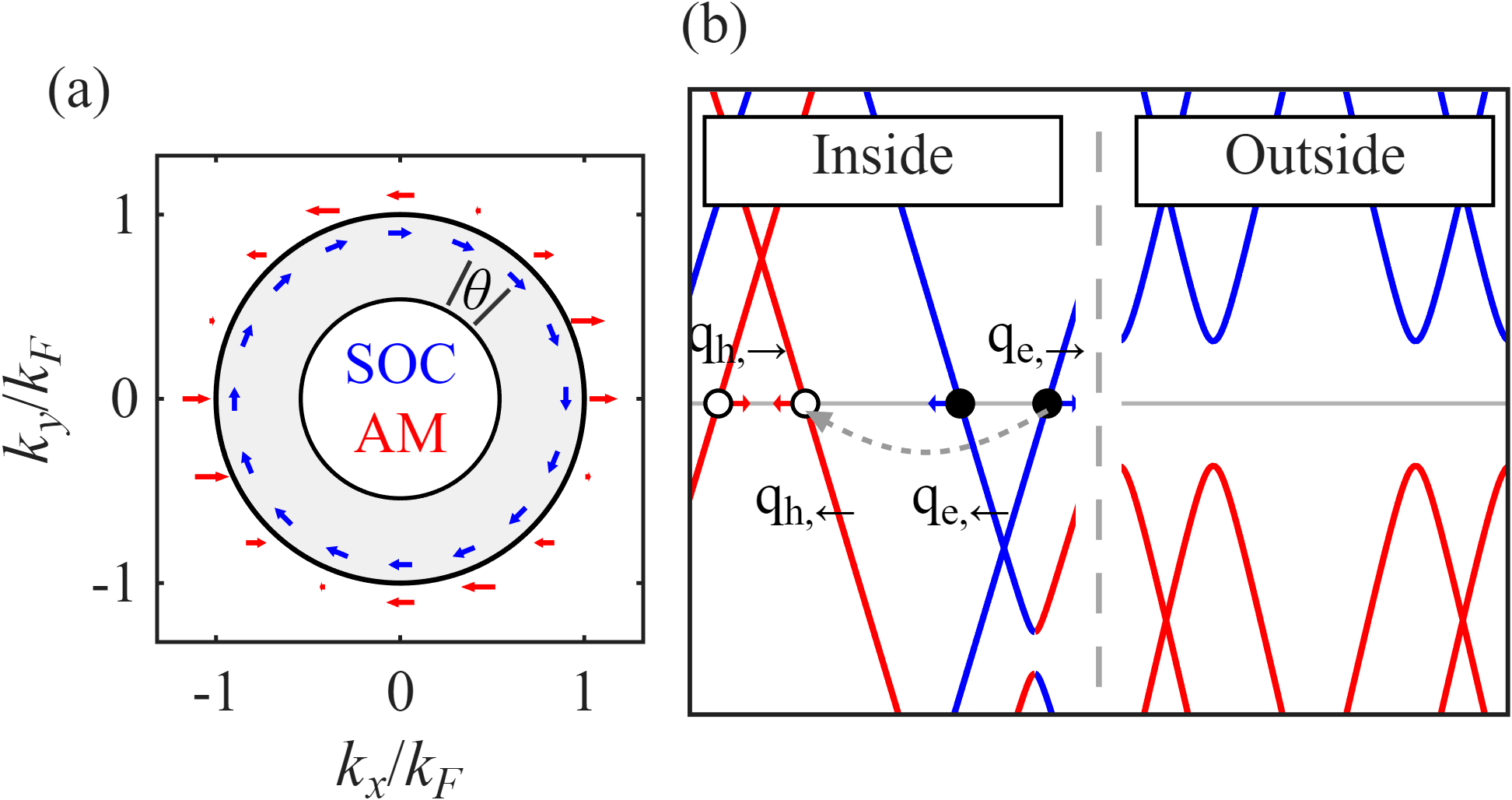}
\caption{Momentum-space structure and Andreev confinement. (a) Helical spin–momentum locking of the TI surface state (blue arrows) and the collinear, momentum-dependent altermagnetic spin texture (red arrows). The angle $\theta$ parametrizes the orientation of the $d$-wave spin-splitting pattern in momentum space. For $\theta=0$, its nodal lines lie along $k_x=\pm k_y$. (b) Schematic quasiparticle spectrum at $k_x=0$. Underneath the strip, altermagnetic splitting produces low-energy propagating Bogoliubov modes, whereas the surrounding superconducting surface remains gapped.}
\label{fig:model}
\end{figure}

We model the helical surface state by the minimal lattice-regularized Hamiltonian
\begin{equation}
    H_0(\bm k)=-\mu\,\tau_3+v\left(\sin k_y\,s_1-\sin k_x\,s_2\tau_3\right),
    \label{eq:H0}
\end{equation}
where $\bm k=(k_x,k_y)$ is the crystal momentum in the plane, $v$ is the Dirac velocity, $\mu$ is the chemical potential, and $s_i$ and $\tau_i$ are Pauli matrices in spin and Nambu spaces, respectively. The corresponding helical spin texture is indicated by the blue arrows in Fig.~\ref{fig:model}(a). As a lattice representation of a surface Dirac theory, it also contains additional Dirac cones at other time-reversal-invariant momenta. In Appendix~\ref{app:singlecone}, we introduce a Wilson-type perturbation that gaps out the additional cones while leaving the target $\bar Y$ cone gapless, and show that their removal produces only perturbative corrections to the gap-closing boundaries near the target $\bar Y$ cone. Underneath the altermagnetic strip this surface state additionally experiences a collinear altermagnetic field along $s_1$, whose $d$-wave momentum dependence is modeled as
\begin{equation}
\begin{aligned}
    J(\bm k)=&2J\left[\cos(2\theta)(\cos k_y-\cos k_x)\right.\\
    &\quad\; -\left.\sin(2\theta)\sin k_x\sin k_y\right].
\end{aligned}
\label{eq:Jglobal}
\end{equation}
Here $J$ is the altermagnetic field strength, and $\theta$ parametrizes the orientation of the $d$-wave momentum-space spin-splitting pattern relative to the strip axes, while the spin axis is fixed along $x$. Unlike the helical field in Eq.~(\ref{eq:H0}), whose spin direction winds with momentum, $J(\bm k)s_1$ remains collinear while changing magnitude and sign across the Brillouin zone, as shown by the red arrows in Fig.~\ref{fig:model}(a). Including superconducting proximity pairing, the BdG Hamiltonians inside and outside the strip are therefore
\begin{equation}
    H(\bm k)=H_0(\bm k)+\begin{cases}
J(\bm k)s_1\tau_3+\Delta' s_2\tau_2, &|y|<W/2,\\
        \Delta s_2\tau_2, &|y|>W/2,
    \end{cases}
\label{eq:Hglobal}
\end{equation}
where $\Delta'$ and $\Delta$ are the superconducting gaps underneath and outside the strip, respectively. For the finite-width geometry, the $k_y$ dependence in Eqs.~(\ref{eq:H0})-(\ref{eq:Hglobal}) is implemented as real-space hopping along $y$. The equations above should therefore be understood as the Bloch Hamiltonians of the two locally uniform regions.

In superconducting TI heterostructures, proximity-induced gaps are typically in the sub-meV to meV range, well below the electronic energy scale $v$ of the Dirac surface state \cite{Xu2014,XuPRL2014}. We assume that the induced altermagnetic exchange amplitude $J$ is similarly small. We further choose the chemical potential $\mu$ sufficiently close to the Dirac point that the Fermi pocket around the target cone occupies only a small patch of the Brillouin zone. Such a regime can be accessed experimentally by electrostatic gating~\cite{Kim2012} or compensation doping. We therefore work in the low-energy regime
\begin{equation}
    \max\left\{|\mu|,4J,\Delta,\Delta'\right\}\ll v
    \label{eq:lowenergyregime}
\end{equation}

For the parameter regime of interest, we focus on the low-energy sector near the Brillouin-zone boundary $\bar Y=(0,\pi)$, where the altermagnetic field takes the value $J(\bar Y)=-4J\cos(2\theta)$. As shown below, this finite field shifts the particle- and hole-like branches in opposite $k_y$ directions and provides the key mechanism for Andreev confinement. Surface Dirac cones centered at non-$\bar\Gamma$ momenta such as $\bar Y$ can arise as genuine surface states in microscopic TI models~\cite{TeoFuKane2008,Roy2014Surface} and have been observed in Zr$_2$Te$_2$P and Ti$_2$Te$_2$P~\cite{Ji2016,Dhakal2022}. Thus, we take the $\bar Y$ cone as a representative physical realization.

To analyze the confinement mechanism near $\bar Y$, we introduce the local momentum $\bm q$ via $\bm k= \bar Y + \bm q =(q_x,\pi+q_y)$. For $|\bm q|\ll1$, the altermagnetic field becomes
\begin{equation}
    J_{\bar Y}(\bm q)=-4J_c+J_c(q_x^2+q_y^2)+2J_s q_xq_y,
\label{eq:Jdecomp}
\end{equation}
where $J_c=J\cos(2\theta)$ and $J_s=J\sin(2\theta)$. 
The low-energy Hamiltonian underneath the strip is
\begin{align}
    H_{\bar Y}(\bm q)=& -\mu\tau_3-v(q_y s_1+q_xs_2\tau_3) \notag \\
     & +J_{\bar Y}(\bm q)s_1\tau_3+\Delta' s_2\tau_2.
    \label{eq:Hmodel}
\end{align}
Outside the strip $J_{\bar Y}=0$ and $\Delta'$ is replaced by $\Delta$.

An Andreev-confined low-energy channel requires propagating subgap quasiparticles underneath the strip while the surrounding fully gapped region supports no propagating states at the same energy. We first consider normal incidence with $q_x=0$, for which $s_1\tau_3$ is conserved. To obtain the leading analytical result, we further neglect the quadratic correction $J_cq_y^2$ and approximate $J_{\bar Y}\simeq-4J_c$. In the eigensector of $s_1\tau_3$ with eigenvalue $\eta=\pm1$, the Hamiltonian inside the strip reduces to
\begin{equation}
    H_{\eta}(q_y)=-4\eta J_c-(\eta vq_y+\mu)\rho_3+\Delta'\rho_1,
    \label{eq:Heta}
\end{equation}
where $\rho_1=s_2\tau_2$ and $\rho_3=\tau_3$ act within the two-component $\eta$ sector. The corresponding spectrum is
$E_{\eta,\pm}(q_y)=-4\eta J_c\pm\sqrt{(\eta vq_y+\mu)^2+|\Delta'|^2}$. The resulting low-energy bands underneath the strip are illustrated schematically in Fig.~\ref{fig:model}(b). The zero-energy condition admits propagating solutions only when
\begin{equation}
    4|J_c|>\Delta',
    \label{eq:gaplesscondition}
\end{equation}
with transverse momenta $q_y=-\frac{\eta}{v}\big(\mu\pm\sqrt{16J_c^2-|\Delta'|^2}\big)$.
Equation~(\ref{eq:lowenergyregime}) ensures that these solutions remain within the local $\bar Y$ expansion.

Outside the strip, the absence of altermagnetic field leaves the fully gapped helical superconductor with spectrum
$E_{\eta,\pm}(q_y)=\pm\sqrt{(\eta vq_y+\mu)^2+\Delta^2}$.
Hence states with $|\epsilon|<\Delta$ are evanescent outside the strip but can propagate across the altermagnetic region when $|\epsilon+4\eta J_c|>\Delta'$, as illustrated in Fig.~\ref{fig:model}(b). 

In the limit $\Delta'=0$, the propagating modes become purely electron- or hole-like. For $\mu>0$ and $q_x=0$, their momenta may be organized according to the propagation direction as
\begin{subequations}
\begin{align}
    q_{e,\rightarrow(\leftarrow)}&=-\frac{4J_c}{v} + (-)\frac{\mu+\epsilon}{v},\\
    q_{h,\leftarrow(\rightarrow)}&=\frac{4J_c}{v}+(-)\frac{\mu-\epsilon}{v}.
\end{align}
\label{eq:Fermipoints}
\end{subequations}
These four propagating states constitute the scattering channels that enter the Andreev-quantization condition.
Below, we use the leading-order theory with $\Delta'=0$ to illustrate the confinement mechanism. Corrections from the lattice dispersion and finite $\Delta'$ are discussed in the Appendices~\ref{app:latticeboundary} and \ref{app:finitepairing}.

\section{Andreev Confinement and Topological Phase Diagram}
\label{sec:andreev}

\begin{figure}
\includegraphics[width=0.45\textwidth]{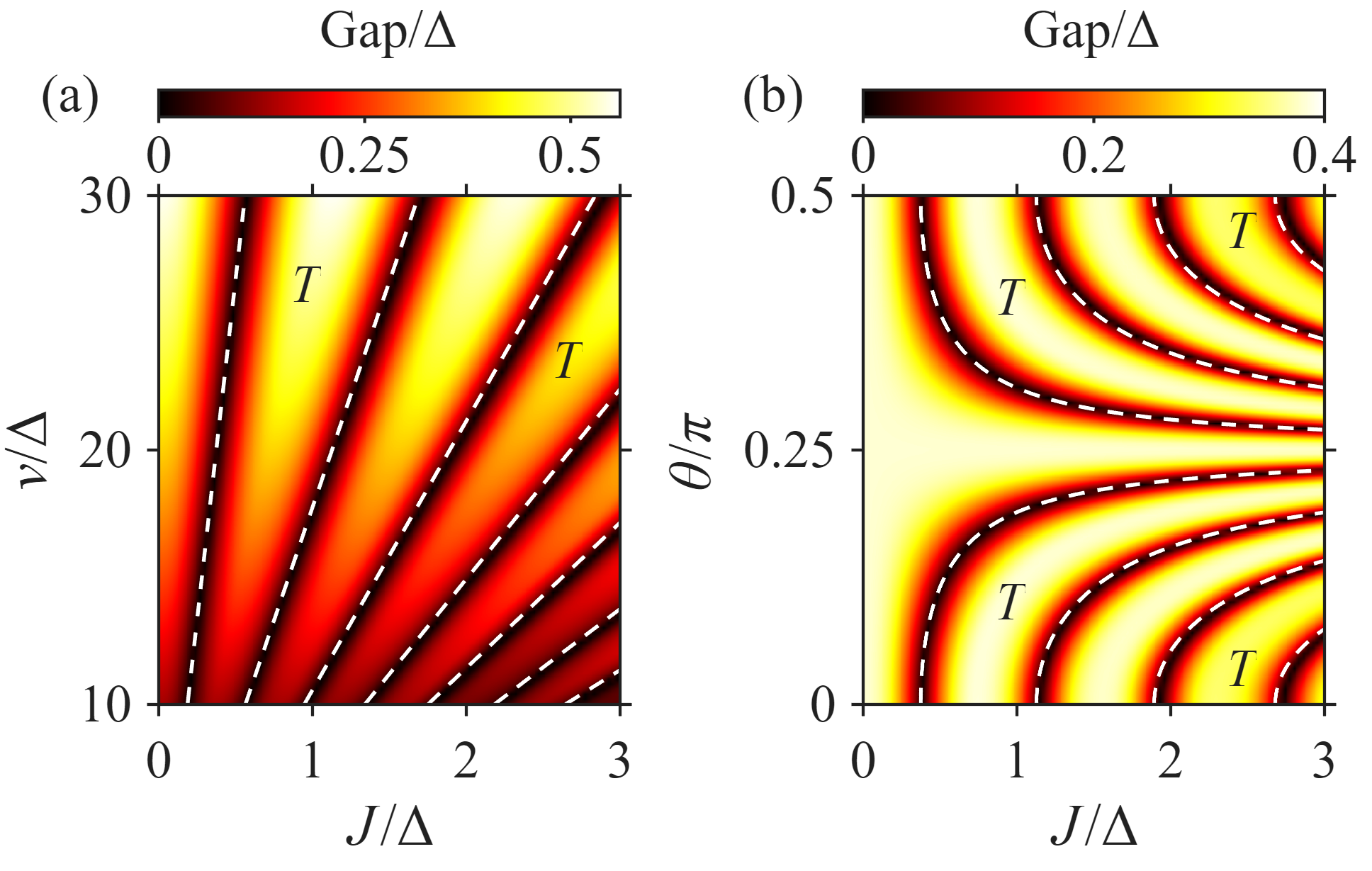}
\caption{Topological phase diagrams of the altermagnetic strip. The color scale shows the quasiparticle gap normalized by $\Delta$. The regions labeled by $T$ are topological, while the remaining regions are trivial. (a) Gap as a function of $J$ and $v$ at $\theta=0$. (b) Gap as a function of $J$ and $\theta$ at $v=1$. White dashed lines indicate the theoretical gap-closing boundaries including lattice corrections (see Appendix~\ref{app:latticeboundary}). Both panels use parameters $\Delta=0.05$, $\Delta'=0$, $\mu=0.1$, $W=21$, and $-100\leq y\leq100$. }
\label{fig:phasediagram}
\end{figure}

We first focus on the sequence of confined-band gap closings at normal incidence, $q_x=0$, which determines the boundaries shown in Fig.~\ref{fig:phasediagram}. At $q_x=0$, the eigenvalues $\eta=\pm1$ of $s_1\tau_3$ are conserved. The scattering problem separates into two decoupled $\eta$ sectors. The propagating strip modes are eigenstates of $\rho_3=\tau_3$ and purely electron- or hole-like. For each $\eta$ sector, the electron- and hole-like modes have opposite transverse group velocities. Thus, an incoming electron has no counterpropagating electron state in the same sector, and similarly for an incoming hole. Because $\eta$ is conserved across the interfaces, normal reflection is absent, leaving only Andreev reflection for subgap states (see Fig.~\ref{fig:model}).

The zero-energy Andreev-reflection follows directly from the evanescent states outside the strip. In a fixed $\eta$ sector, the superconductor is described by
\begin{equation}
    H_{\eta}=-(\eta v q_y+\mu)\rho_3+\Delta\rho_1 .
    \label{eq:SCeta}
\end{equation}
At zero energy ($\epsilon = 0$), the allowed complex transverse momenta satisfy $\eta v q_y+\mu=\mp i\Delta$. 
Plugging this into the BdG equation gives
\begin{equation}
\rho_2\psi_\eta=\mp\psi_\eta ,
\label{eq:rho2decay}
\end{equation}
where $\rho_2=-i\rho_3\rho_1$ completes the Pauli algebra in the reduced BdG space. The decaying solutions on the two sides of the strip have opposite $\rho_2$ eigenvalues. In the electron-hole basis, these eigenstates are proportional to $(1,\mp i)^T$. The Andreev reflection introduces a relative phase of $\pm\pi/2$. The scattering matrices at the upper and lower interfaces can be written as
\begin{equation}
    S_+=\begin{pmatrix}
    0&-i\\
    i&0
    \end{pmatrix},
    \quad
    S_-=\begin{pmatrix}
    0&i\\
    -i&0
    \end{pmatrix}.
    \label{eq:Smatrix}
\end{equation}

\begin{figure}
\includegraphics[width=0.45\textwidth]{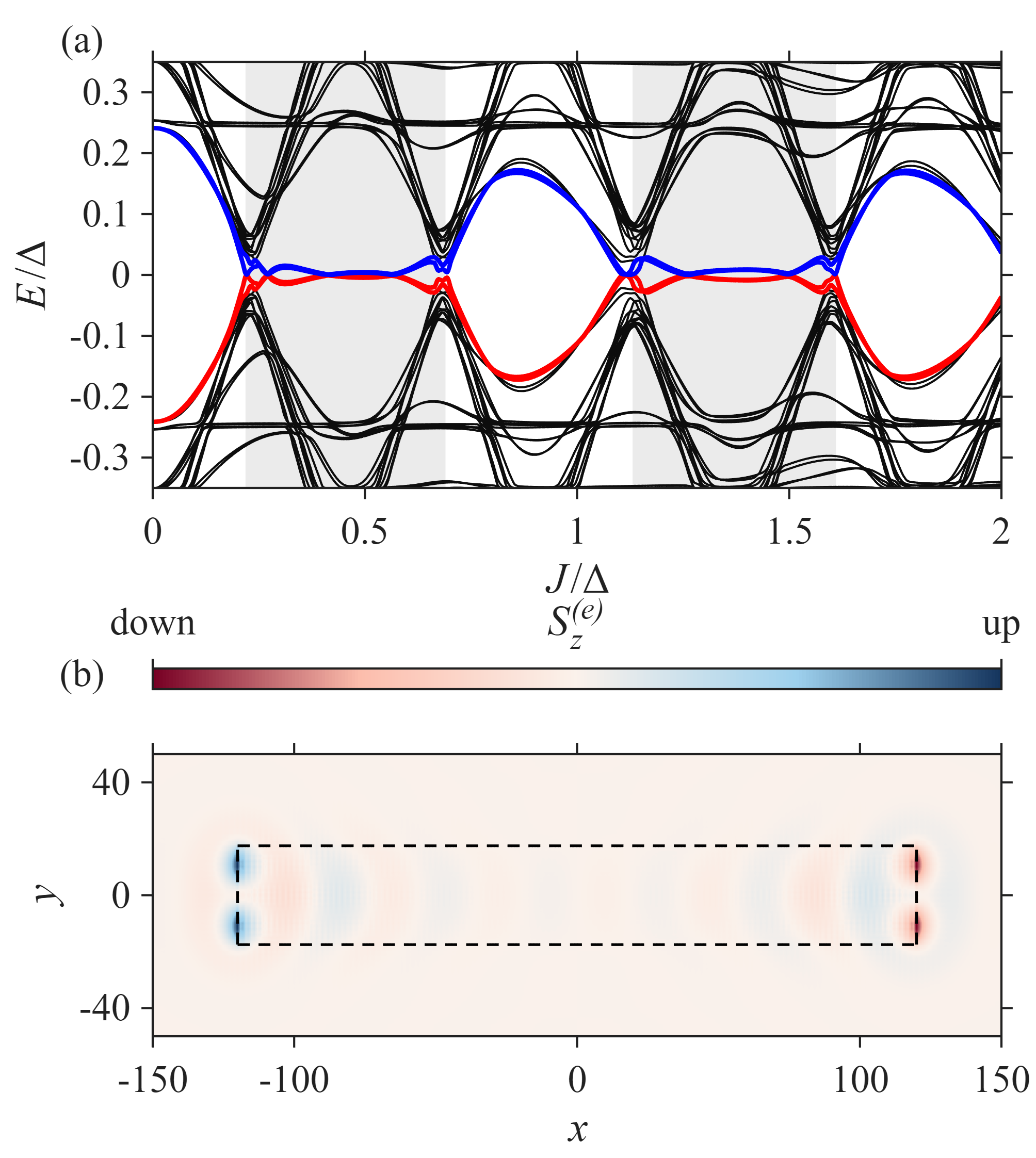}
\caption{Finite-strip spectrum and spin polarization of the Majorana end states for $\Delta=0.05$, $\Delta'=0$, $\mu=0.1$, $v=1$, $L=241$, $W=35$, $-200\leq x\leq 200$, and $-50\leq y\leq 50$. (a) Energy spectrum of the finite strip. The black bands denote the bulk states, while the blue and red bands denote the electron- and hole-like end modes, respectively. The alternating band inversions indicate successive topological regions. (b) Spin-resolved electronic density profile of the MZMs at $J=1.3\Delta$, showing a double-peak structure at each end. 
}
\label{fig:zeromodes}
\end{figure}

Propagation between the two strip boundaries is described by the diagonal phase matrices 
\begin{subequations}
\label{eq:Tmatrix}
\begin{align}
T_+&=\mathrm{diag}\left(e^{iq_{e,\rightarrow}W},e^{iq_{h,\rightarrow}W}\right),\\
    T_-&=\mathrm{diag}\left(e^{-iq_{e,\leftarrow}W},e^{-iq_{h,\leftarrow}W}\right),
\end{align}
\end{subequations}
where the transverse momenta are given by Eq.~(\ref{eq:Fermipoints}). Starting with an upward-moving state at the lower boundary, one complete round trip is represented by $S_-T_-S_+T_+$. A bound state forms when this round-trip operator has eigenvalue one, or equivalently
$\det\big(\mathbb{I}-S_-T_-S_+T_+\big)=0$, where $\mathbb{I}$ denotes the identity matrix. Using Eqs.~(\ref{eq:Smatrix}) and (\ref{eq:Tmatrix}), we obtain
\begin{equation}
    \big[1+e^{-i(q_{h,\leftarrow}-q_{e,\rightarrow})W}\big] \big[1+e^{i(q_{h,\rightarrow}-q_{e,\leftarrow})W}\big] = 0.
    \label{eq:scatteringcondition2}
\end{equation}
The two factors correspond to the two $\eta$ sectors. At $\epsilon=0$, Eq.~(\ref{eq:Fermipoints}) gives $\Delta q=q_{h,\leftarrow}-q_{e,\rightarrow}=q_{h,\rightarrow}-q_{e,\leftarrow}={8J_c}/{v}$. 
The chemical potential cancels from the momentum difference $\Delta q$. The zero-energy condition becomes
\begin{equation}
\frac{|J\cos(2\theta)|}{v}=\frac{(n+1/2)\pi}{4W},\quad n\in\mathbb{N}.
\label{eq:approx}
\end{equation}
This equation gives the successive values of $J$ at which confined quasiparticle branches cross zero energy. The finite-strip spectrum in Fig.~\ref{fig:zeromodes}(a) shows end modes in alternating parameter regions, consistent with the sequence of trivial and topological phases shown in Fig.~\ref{fig:phasediagram}(a). A distinctive feature is that the effective field strength is $J_c=J\cos(2\theta)$. Thus, rotating the altermagnetic texture shifts the same sequence of transitions to larger $J$ as $|\cos(2\theta)|$ decreases, giving the characteristic bending of the phase boundaries in Fig.~\ref{fig:phasediagram}(b). This angular tunability also makes the condition $4J\ll v$ in Eq.~(\ref{eq:lowenergyregime}) somewhat conservative for the normal-incidence problem, since the normal-incidence low-energy physics is controlled primarily by $J\cos(2\theta)$.

Near the $n$-th phase boundary, the two particle-hole-related confined levels at $q_x=0$ are
\begin{equation}
    \epsilon_{\pm}=\pm\frac{4W}{W+\xi}\left[|J\cos(2\theta)|-\frac{v}{4W}\left(n+\frac12\right)\pi\right],
    \label{eq:gapclosingdispersion}
\end{equation}
where $\xi=v/\Delta$ defines the superconducting coherence length. The two branches cross linearly at $\epsilon=0$ around $|J\cos(2\theta)|=v/(4W)\left(n+1/2\right)\pi$, explicitly demonstrating the closing and reopening of the confined quasiparticle gap across each phase boundary.
The topological regions are given by $4W|J\cos(2\theta)|/(\pi v)\in(2m+1/2,2m+3/2)$ with integer $m=0,1,\ldots$,
%\begin{equation}
%    \frac{(2m+1/2)\pi}{4W}<\frac{|J\cos(2\theta)|}{v}<\frac{(2m+3/2)\pi}{4W},\,m\in\mathbb{N},
%\end{equation}
or, more compactly 
\begin{equation}
    \cos\left(\frac{4WJ\cos(2\theta)}{v}\right)<0.
\end{equation}
Within the low-energy description, these alternating phases can thus be characterized by a $\mathbb{Z}_2$ invariant
\begin{equation}
(-1)^\nu=\mathrm{sgn}\left(\cos\left(\frac{4WJ\cos(2\theta)}{v}\right)\right),\quad \nu\in\mathbb{Z}_2,
\end{equation}
where $\nu=0$ and $\nu=1$ denote the trivial and topological phases, respectively.

\section{Majorana End States and Spin Polarization}
\label{sec:majorana}

\subsection{Low-energy effective model of the Andreev bands}
\label{sec:andreev-effective}
We next consider a finite strip in the topological regime and derive the Majorana states localized at its ends. We begin with the confined spectrum at $q_x=0$. Repeating the Andreev-quantization analysis of Sec.~\ref{sec:andreev} for finite energies $|\epsilon|<\Delta$, while retaining $\Delta'=0$ and $J_{\bar Y}\simeq-4J_c$, gives the quantization equation
\begin{equation}
    ({W}/{v})\left(\epsilon+4\eta J_c\right)-\arccos\left({\epsilon}/{\Delta}\right)=n\pi,\quad n\in\mathbb Z.
    \label{eq:finiteEnergyQuantization}
\end{equation}
The two $\eta$ sectors are related by particle-hole symmetry. Thus, the solutions occur in pairs with opposite energies. Denoting the subgap solutions by $\epsilon_{\eta,n}$, the excitation gap at $q_x=0$ is given by $\varepsilon=\min_{n}\left|\epsilon_{\eta,n}\right|$, where the minimum runs over the subgap solutions. At a gap-closing phase boundary $\varepsilon=0$, the finite-energy quantization Eq.~(\ref{eq:finiteEnergyQuantization}) reduces to Eq.~(\ref{eq:approx}) derived above. 

We focus on the lowest particle-hole-related pair, with energies $\pm\varepsilon$. We assume that this pair is well separated from higher subbands so that the low-energy longitudinal dynamics can be described within the corresponding two-dimensional subspace. To derive its longitudinal dispersion, we retain $q_x$ as a good quantum number and express the transverse coordinate in real space through the substitution $q_y\rightarrow -i\partial_y$. We decompose the BdG Hamiltonian as 
$H(q_x,\partial_y)=H(0,\partial_y)-vq_xs_2\tau_3$
with
\begin{equation}
\begin{aligned}
    &H(0,\partial_y)\\
    =&-\mu\tau_3+i v s_1\partial_y+\begin{cases}
    -4J_c s_1\tau_3,& |y|<W/2,\\
    \Delta s_2\tau_2,& |y|>W/2.
    \end{cases}
\end{aligned}
\label{eq:Htransverse}
\end{equation}
It obeys particle-hole symmetry $ \mathcal C H(q_x,\partial_y)\mathcal C^{-1}=- H(-q_x,\partial_y)$, where $\mathcal C=\mathcal K\tau_1$ and $\mathcal K$ denotes complex conjugation. 

At $q_x=0$, the operator $s_1\tau_3$ is conserved, and particle-hole symmetry $\mathcal C$ connects the two sectors labeled by its eigenvalues $\eta=\pm1$. 
We denote the two lowest particle-hole-related states by $\Phi_\pm(y)$, satisfying
\begin{equation}
  H(0,\partial_y)\Phi_\pm(y)=\pm\varepsilon\Phi_\pm(y).
    \label{eq:PhiDefinition}
\end{equation}
Following the same Andreev-matching procedure as above, $\Phi_+(y)$ with energy $+\varepsilon$ can be written in the basis $\left(\psi_{\uparrow e},\psi_{\downarrow e},\psi_{\uparrow h},\psi_{\downarrow h}\right)^T$ as
\begin{equation}
\Phi_+(y)=\mathcal N e^{-i\eta\mu y/v}F_\eta(y),
\label{eq:PhiPlusExplicit}
\end{equation}
where
\begin{equation}
F_\eta(y)=
\begin{cases}
e^{\kappa(y+W/2)}
\chi_\eta\!\left(\frac{\eta\alpha}{2}\right),
& y<- W/2,\\[6pt]
\chi_\eta(\beta-Qy),
& |y|<W/2,\\[6pt]
e^{2i\beta-\kappa(y-W/2)}
\chi_\eta\!\left(-\frac{\eta\alpha}{2}\right),
& y> W/2.
\end{cases}
\end{equation}
Here, we have defined for convenience the spinor $\chi_\eta(\varphi)= (e^{i\varphi},
\eta e^{i\varphi}, \eta e^{-i\varphi},
-e^{-i\varphi})^T$, and quantities $\alpha=\arccos\left({\varepsilon}/{\Delta}\right)$, $\kappa={\sqrt{\Delta^2-\varepsilon^2}}/{v}$, $Q=(4J_c+\eta \varepsilon)/{v}$, and $\beta={(\eta\alpha-QW)}/{2}$. 
The normalization factor is $\mathcal N={1}/{2\sqrt{W+\kappa^{-1}}}$.  The negative-energy partner follows from particle–hole symmetry, i.e.,
\begin{equation}
    \Phi_-(y)=\mathcal C\Phi_+(y)=\tau_1\Phi_+^*(y).
    \label{eq:PhiMinusExplicit}
\end{equation}

Projecting $H(q_x,\partial_y)$ onto the subspace spanned by $\{\Phi_+,\Phi_-\}$ yields the effective Hamiltonian  
\begin{equation}
H_{\rm eff}(q_x) =\varepsilon\sigma_3+uq_x\sigma_2,
\label{eq:projectedMatrix}
\end{equation}
where $\sigma_i$ are Pauli matrices acting in the projected subspace.
This Hamiltonian has the form of a 1D massive Dirac Hamiltonian, with effective mass $\varepsilon$ and velocity
\begin{equation}
\begin{aligned}
    u=&\frac{v\kappa [q_0 \cos(q_0 W)+{\kappa}\sin(q_0 W)]}{q_0(W+\kappa^{-1})(\kappa^2+q_0^2)},\quad q_0\equiv \frac{\mu}{v}.
\end{aligned}
\label{eq:uProjection}
\end{equation}
In the limit $\mu\to0$, this velocity approaches $v$. The lowest-energy pair thus disperses as $E_{\pm}(q_x)=\pm\sqrt{\varepsilon^2+u^2q_x^2}$.
The gap at $q_x=0$ closes when $\varepsilon=0$, reproducing the transition points obtained from the Andreev quantization condition. 

%{\color{red}We now consider an altermagnetic strip of finite length $L$. Within the same leading approximation used above, with $\Delta'=0$ and $J_{\bar Y}\simeq-4J_c$ underneath the strip, the continuum Hamiltonian is
%\begin{equation}
%    \begin{aligned}
%    H(q_x,\partial_y)={}&-\mu\tau_3+i v s_1\partial_y- v s_2\tau_3q_x\\
%    &+\begin{cases}
%    -4J_c s_1\tau_3,& |x|<L/2,\ |y|<W/2,\\[4pt]
%    \Delta s_2\tau_2,& \text{otherwise}.
%    \end{cases}
%    \end{aligned}
%    \label{eq:HfiniteStrip}
%\end{equation}
%Thus the finite altermagnetic region terminates at $x=\pm L/2$, while beyond either longitudinal end the surface is the uniformly proximitized helical superconductor with no altermagnetic field.}

\subsection{Asymptotic Majorana wave functions}
We now consider a strip of finite length $L$, terminated at $x=\pm L/2$, in the topological regime. We use the lowest confined pair to determine the asymptotic Majorana wave functions inside the strip, neglecting hybridization between the two ends~\footnote{Unfortunately, this two-state effective theory is insufficient for exact wave-function matching between the regions inside and outside the strip at its ends, whereas including higher bands in the full solution would not lead to a compact analytical result.}. Replacing $q_x$ by $-i\partial_x$ in Eq.~(\ref{eq:projectedMatrix}), the zero-energy equation for either end reads
\begin{equation}
    \left(-iu\sigma_2\partial_x+\varepsilon\sigma_3\right)\Psi_{R(L)}(x)=0.
    \label{eq:MajoranaInteriorDirac}
\end{equation}
Selecting solutions that decay into the strip from the left and right ends gives
\begin{subequations}
\begin{align}
& \Psi_R(x) \propto e^{-(L/2-x)/\ell}\chi_{\mathrm{sgn}(u)}
\label{eq:MajoranaInteriorRight} \\
& \Psi_L(x) \propto e^{-(L/2+x)/\ell}\chi_{-\mathrm{sgn}(u)}
\label{eq:MajoranaInteriorLeft}
\end{align}
\end{subequations}
where $\ell={|u|}/{\varepsilon}$ and $\chi_{s}$ is an eigenstate of $\sigma_1$ satisfying $\sigma_1\chi_{s}=s\chi_{s}$, $s=\pm1$. 
For $L\gg\ell$, the overlap between the two end modes
is exponentially small. A convenient phase convention for the two $\sigma_1$ eigenvectors is $ \chi_s= e^{i\pi(1-s)/4} ( 1,  s )^T.$
Restoring the transverse wave functions, the wavefunctions of the end modes read
\begin{subequations}
\label{eq:Majorana}
\begin{align}
    \Psi_L(x,y)=&\mathcal{N}_M e^{-(L/2+x)/\ell+i\pi(1+\mathrm{sgn}(u))/4} \notag \\
    &\times[\Phi_+(y)-\mathrm{sgn}(u)\Phi_-(y)],\\
    \Psi_R(x,y)=& \mathcal{N}_M e^{-(L/2-x)/\ell+i\pi(1-\mathrm{sgn}(u))/4} \notag \\
    &\times [\Phi_+(y)+\mathrm{sgn}(u)\Phi_-(y)], \label{eq:MajoranaTransverseSpinors}
\end{align} 
\end{subequations}
where $\mathcal{N}_M$ is a real normalization factor. 
Since $\Phi_-=\mathcal C\Phi_+$ and $\mathcal C^2=1$, this phase convention ensures the Majorana self-conjugacy condition $\mathcal C\Psi_{L/R}=\Psi_{L/R}$. The leading approximation captures the longitudinal localization of the MZMs at the strip ends, consistent with the full lattice calculation shown in Fig.~\ref{fig:zeromodes}(b). Interestingly, the numerical spin density also exhibits a transverse multi-peaked profile at each end~\footnote{The numerical spin density is averaged over the four in-gap eigenstates closest to zero energy to obtain a basis-independent profile of the near-zero-energy subspace.}. The detailed transverse profile observed at larger $|J_c|/v$ is not reproduced by the leading approximation. We examine its formation through wave-function matching at the strip ends below.

\begin{figure}
\includegraphics[width=0.45\textwidth]{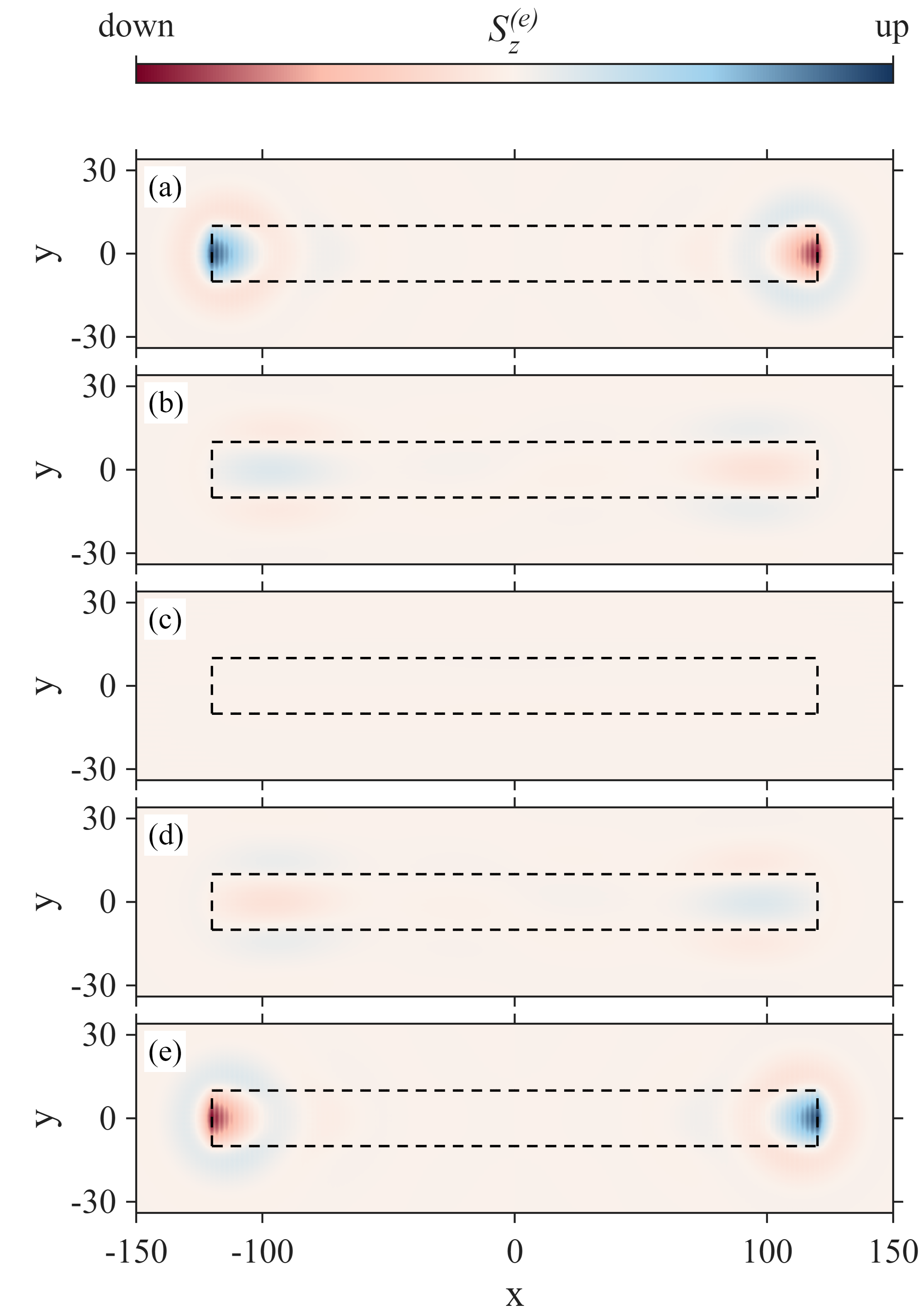}
\caption{Evolution of the Majorana end-state electronic spin distribution with the orientation angle $\theta$ of the $d$-wave spin-splitting pattern. Panels (a)-(e) correspond to $\theta=0$, $\pi/8$, $\pi/4$, $3\pi/8$, and $\pi/2$, respectively. The end states exhibit single-peaked transverse spin-density profiles and opposite spin polarizations in the two orientation ranges separated by the trivial regime. No localized MZM is present at $\theta=\pi/4$. Parameters: $\Delta=0.05$, $\Delta'=0$, $\mu=0.1$, $v=1$, $J=0.55\Delta$, $L=241$, $W=21$, $-200\leq x\leq 200$, and $-50\leq y\leq 50$. At $\theta=0$, the strip lies in the first topological region, $0.36\lesssim J/\Delta\lesssim1.10$.}
\label{fig:thetaevolution}
\end{figure}

Before examining the transverse profile, we use $\ell(\theta)=|u(\theta)|/\varepsilon(\theta)$ to determine how the longitudinal localization depends on the altermagnetic orientation. Using the quantization condition in Eq.~(\ref{eq:finiteEnergyQuantization}) and the velocity in Eq.~(\ref{eq:uProjection}), the localization length takes simple forms in two limits:
\begin{equation}
\begin{aligned}
    \ell(\theta)
    &\simeq
    \begin{cases}
    \frac{\xi}{d(\theta)}
    \frac{\left|\cos(q_0 W)+\frac{\Delta}{\mu}\sin( q_0W)\right|}{1+\left(\frac{\mu}{\Delta}\right)^2},
    & W\gg\xi,\\[8pt]
    \frac{\xi}{\sin d(\theta)}
    \frac{\cos^2 d(\theta)|\cos(q_0 W)|}{\cos^2 d(\theta)+\left(\frac{\mu}{\Delta}\right)^2},
    & W\ll\xi,
    \end{cases}
\end{aligned}
\label{eq:MajoranaLocalizationTheta}
\end{equation}
where $\xi=v/\Delta$, $q_0=\mu/v$, and
\begin{equation}
\begin{aligned}
    d(\theta)
    =&\,\arcsin\left|\cos\left({4WJ\cos(2\theta)}/{v}\right)\right|
\end{aligned}
\label{eq:dTheta}
\end{equation}
measures the distance of the quantization phase from its nearest gap-closing value. These results describe the localization length within the topological regions. As $\theta$ approaches a topological phase boundary, $d(\theta)\to0$ and the confined-band gap closes. Provided that the effective velocity remains nonzero, the localization length diverges as $\ell\propto1/d(\theta)$. The MZMs therefore extend farther into the strip near a transition and increasingly overlap when $\ell$ becomes comparable to $L$. Away from the phase boundary, the reopened gap allows the end modes to become more localized. This behavior is consistent with the numerical spin distributions in Fig.~\ref{fig:thetaevolution}. The end modes become more extended along the strip as $\theta$ increases from $0$ to $\pi/8$, and no localized MZM is present in the trivial regime at $\theta=\pi/4$. Upon further rotation, the system reenters the topological regime and localized end modes reappear.

\subsection{Spin structure of the Majorana end modes}
Using the Majorana wave functions in Eqs.~\eqref{eq:Majorana}, we calculate the local probability density and spin density,
\begin{subequations}
\label{eq:electronSpinDensity}
\begin{align}
     n_{L/R}(x,y)= & \Psi_{L/R}^\dagger(x,y)P_e\Psi_{L/R}(x,y), \\
    \bm S_{L/R}(x,y)=& \frac12\Psi_{L/R}^\dagger(x,y)P_e\bm s\Psi_{L/R}(x,y), 
\end{align}
\end{subequations}
where $\bm s=(s_1,s_2,s_3)$ is the vector of spin Pauli matrices and $P_e=(1+\tau_3)/2$ is the projector onto electron sector. The electronic spin density characterizes the spin-resolved spectral weight relevant to spin-polarized tunneling measurements~\cite{Sticlet2012}.

Substituting $\Phi_\pm$ from Eqs.~\eqref{eq:PhiPlusExplicit} and \eqref{eq:PhiMinusExplicit} into Eq.~(\ref{eq:Majorana}), we find, inside the strip ($|x|<L/2$ and $|y|<W/2$),
\begin{subequations}
\begin{align}
    n_{L/R}(x,y) & =n_0e^{-2(L/2\pm x)/\ell},
    \label{eq:density} \\
    \bm S_{L/R}(x,y)&=\mp\frac{n_{L/R}}{2}\,\mathrm{sgn}(u)\eta \notag\\
    &\times\left(0,-\sin(2q_0 y),\cos(2 q_0 y)\right).
    \label{eq:spinPolarizationGeneral}
\end{align}
\end{subequations}
Here, the upper and lower signs refer to the left and right ends, respectively, and $n_0=4\mathcal N^2\mathcal N_M^2$. Within the same leading approximation, the probability densities are related by $n_L(x,y)=n_R(-x,y)$, while the spin densities satisfy
\begin{equation}
    \bm S_L(x,y)=-\bm S_R(-x,y).
\label{eq:endSpinPolarizationRelation}
\end{equation}
The finite-$\Delta'$ correction to the transverse electronic density of an individual confined state is discussed in Appendix~\ref{app:finitepairingprofile}.

The dependence on $\theta$ follows from the transformation properties of the Hamiltonian in Eq.~(\ref{eq:Hglobal}). Under time-reversal $\mathcal{T}=is_2\mathcal{K}$, $ \mathcal{T}H(\partial_x,\partial_y;\theta)\mathcal{T}^{-1}=H\left(\partial_x,\partial_y;\theta+{\pi}/{2}\right)$, while under mirror reflection $\mathcal{M}_x=is_1\tau_3\mathcal{R}_x$, $\mathcal{M}_xH(\partial_x,\partial_y;\theta)\mathcal{M}_x^{-1}=H(\partial_x,\partial_y;-\theta)$, where $\mathcal{R}_x$ denotes spatial reflection along $x$~\footnote{Since spin $\bm s$ is an axial vector, $\mathcal{M}_x(s_1,s_2,s_3)\mathcal{M}_x^{-1}=(s_1,-s_2,-s_3)$.}.
%These symmetry relations imply that $\Psi\left(x,y;\theta+{\pi}/{2}\right)=\mathcal{T}\Psi(x,y;\theta)$ and $\Psi\left(x,y;-\theta\right)=\mathcal{M}_x\Psi(x,y;\theta)$.
Thus, time reversal flips the spin without exchanging the ends, giving
\begin{align}
    &\bm S_{L/R}\left(x,y;\theta+{\pi}/{2}\right)=-\bm S_{L/R}(x,y;\theta).
\label{eq:spinTextureReversal}
\end{align}
Mirror reflection exchanges the ends and transforms the spin
as $(s_1,s_2,s_3)\rightarrow(s_1,-s_2,-s_3)$. Within the leading approximation, combining this transformation with Eq.~\eqref{eq:endSpinPolarizationRelation} and $S_x=0$ gives
\begin{equation}
    \bm S_{L/R}(x,y;-\theta)=\bm S_{L/R}(x,y;\theta).
    \label{eq:spinTextureMirror}
\end{equation}
These results are consistent with the numerical spin distributions in Fig.~\ref{fig:thetaevolution}. After the intervening trivial regime, localized Majorana end modes reappear with reversed spin polarization at each end, while the two ends remain oppositely polarized.

\subsection{Transverse peak structure from mode interference}
\label{sec:doublepeak}

To resolve the transverse structure near a strip end, we include the higher states omitted from the projection in Sec.~\ref{sec:andreev-effective}. We retain $J_{\bar Y}\simeq-4J_c$ and $\Delta'=0$. We first consider the transverse eigenvalue problem at $q_x=0$, $H_\eta(-i\partial_y)\psi_\eta(y)=\epsilon\psi_\eta(y)$, with $H_\eta$ given by Eq.~\eqref{eq:Heta}. Integrating this equation inside the strip $|y|<W/2$ gives
\begin{equation}
    \psi_\eta(y)=e^{-i\eta\mu(y+W/2)/v}e^{-i\eta(\epsilon+4\eta J_c)(y+W/2)\rho_3/v}\psi_\eta(-W/2).
\end{equation}
For $|y|>W/2$, solving Eq.~\eqref{eq:SCeta} gives
\begin{equation}
    \psi_\eta(y)\propto e^{iq_y y}
    \begin{pmatrix}
    1, 
    (\epsilon+\eta vq_y+\mu)/\Delta
    \end{pmatrix}^T,
\end{equation}
where $q_y=\left(-\eta\mu\pm\sqrt{\epsilon^2-\Delta^2}\right)/v$. The subgap solutions satisfy Eq.~\eqref{eq:finiteEnergyQuantization}, while $|\epsilon|>\Delta$ gives the transverse scattering continuum.

\begin{figure}
\includegraphics[width=0.48\textwidth]{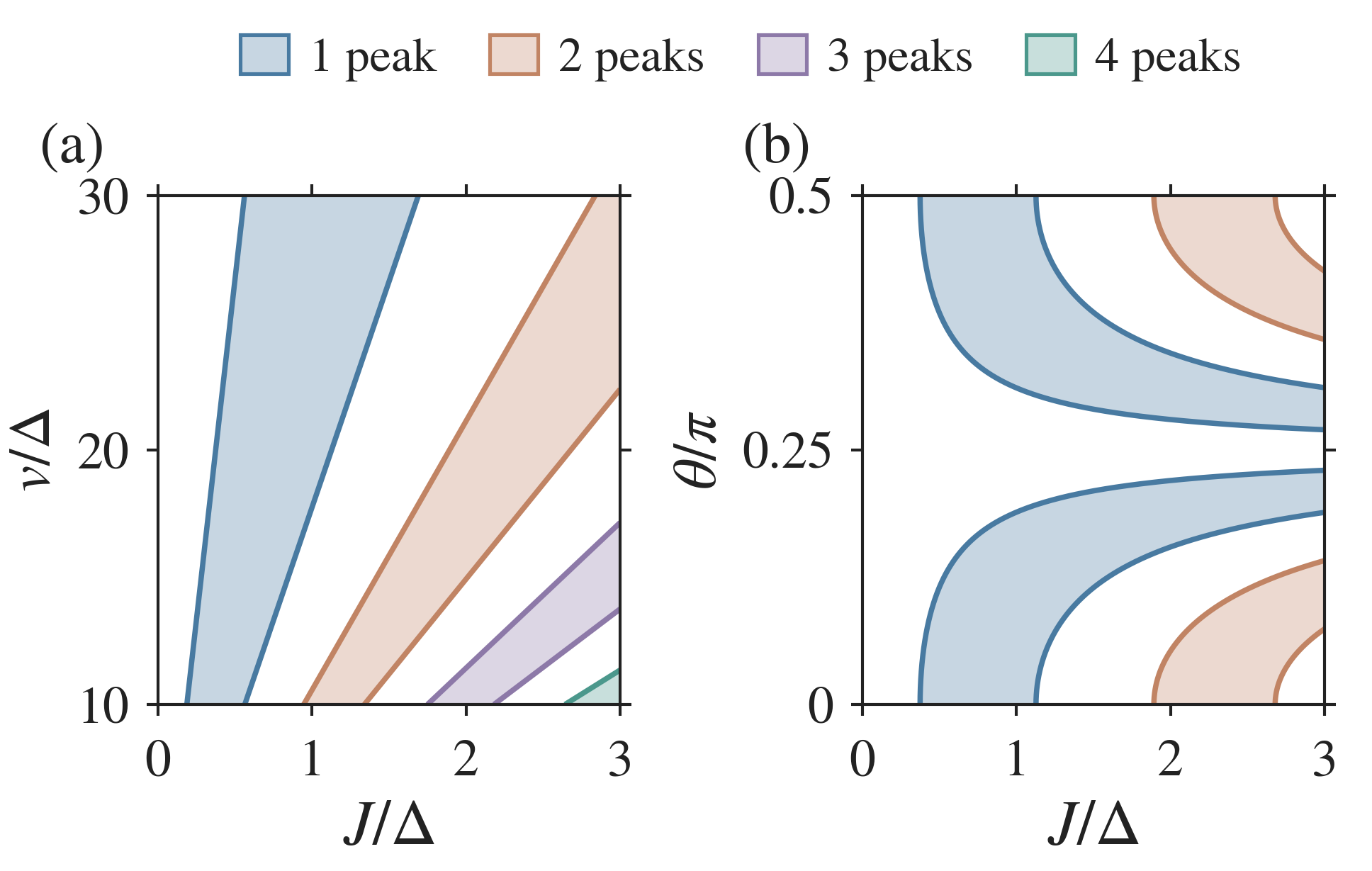}
\caption{Evolution of the transverse peak structure of the Majorana end states across successive topological regions. The schematic phase diagrams are shown as functions of (a) $J/\Delta$ and $v/\Delta$ at $\theta=0$ and (b) $J/\Delta$ and $\theta/\pi$ at $v=1$. The blue, orange, purple, and green regions contain representative end-state profiles with one, two, three, and four peaks at each strip end, respectively. Both panels use parameters $\Delta=0.05$, $\Delta'=0$, $\mu=0.1$, and $W=21$.}
\label{fig:peakregions}
\end{figure}

We next turn to the strip ends at $x=\pm L/2$, where the confined and continuum transverse states are combined to satisfy the end boundary conditions. Restoring the term $-vq_xs_2\tau_3$ couples these transverse components, and the wave function inside $|x|<L/2$ satisfies
\begin{equation}
    \left[H(0,\partial_y)+iv s_2\tau_3\partial_x\right]\Psi_{L/R}(x,y)=0,
    \label{eq:fullEndStateZeroEnergy}
\end{equation}
where $H(0,\partial_y)$ is given by Eq.~\eqref{eq:Htransverse}. Beyond the strip ends ($|x|>L/2$), we expand the solution in Fourier components along $y$. For each real $q_y$, the zero-energy waves with complex longitudinal momentum $q_x$ are proportional to
\begin{equation}
    e^{i(q_xx+q_yy)}\begin{pmatrix}
    \mu\pm i\Delta\\
    -v(q_y+iq_x)\\
    \pm iv(q_y+iq_x)\\
    \pm i(\mu\pm i\Delta)
    \end{pmatrix},
    \label{eq:exteriorZeroEnergyWaves}
\end{equation}
where $v^2(q_x^2+q_y^2)=(\mu\pm i\Delta)^2$. Decay into the exterior requires $\mathrm{Im}\;q_x>0$ on the right and $\mathrm{Im}\;q_x<0$ on the left. At the strip ends, the interior and exterior solutions obey
\begin{subequations}
    \begin{align}
    \Psi_R(L/2-0,y)&=\Psi_R(L/2+0,y),\\
    \Psi_L(-L/2+0,y)&=\Psi_L(-L/2-0,y).
    \end{align}
    \label{eq:endStateContinuity}
\end{subequations}
Substituting the full transverse expansion satisfying Eq.~\eqref{eq:fullEndStateZeroEnergy} and the exterior decaying waves in Eq.~\eqref{eq:exteriorZeroEnergyWaves} into Eq.~\eqref{eq:endStateContinuity} gives linear matching equations for their amplitudes. Requiring all four components to agree for every $y$ determines the allowed coherent combinations of confined and continuum components.

The interference between these components produces the transverse structure of the end state. Figure~\ref{fig:peakregions} summarizes the correspondence between the transverse peak structure and the topological regions. The first four topological regions contain representative end states with one, two, three, and four transverse peaks, respectively. At fixed $W$ and $\theta$, increasing $J/v$ advances the quantization phase $4|J_c|W/v$ through the successive band inversions described by Eq.~\eqref{eq:approx}.

\begin{figure}
\includegraphics[width=0.44\textwidth]{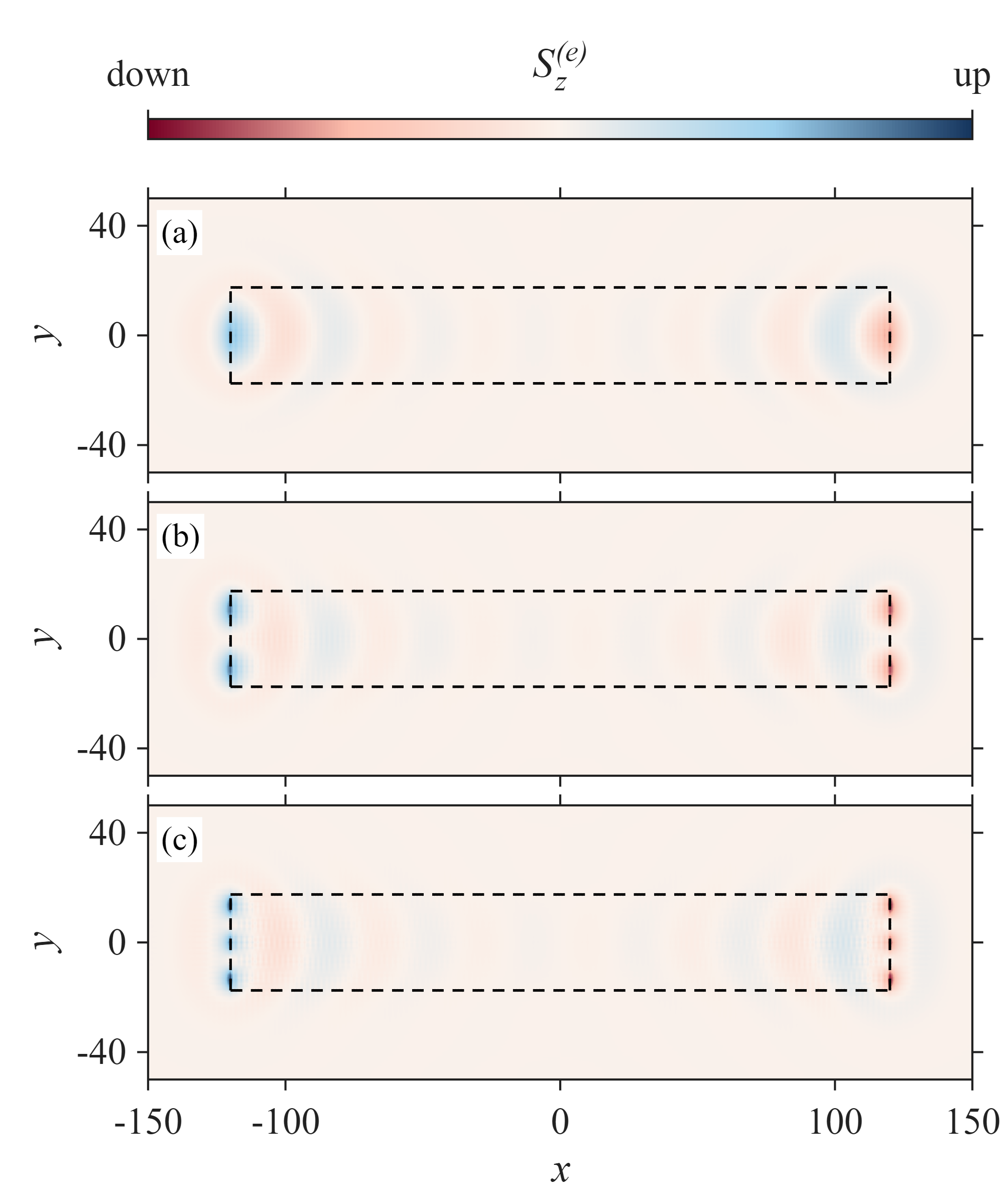}
\caption{Evolution of the Majorana end-state electronic spin distribution with the altermagnetic strength $J$ for $\Delta=0.05$, $\Delta'=0$, $\mu=0.1$, $v=1$, $\theta=0$, $L=241$, $W=35$, $-200\leq x\leq200$, and $-50\leq y\leq50$. Panels (a)-(c) correspond to $J/\Delta=0.45$, $1.3$, and $2.3$ in the first three topological regions, respectively. Increasing $J$ changes the transverse profile of $|S_z^{(e)}|$ from one peak to two and then three peaks at each end, while the two ends remain oppositely polarized.}
\label{fig:peakspin}
\end{figure}

The real-space distributions in Fig.~\ref{fig:peakspin} illustrate this evolution at a fixed width $W=35$. Increasing $J/\Delta$ from $0.45$ to $1.3$ and $2.3$ places the strip in the first, second, and third topological regions, respectively. At each end, the transverse profile changes from a central maximum to two maxima near $y=\pm10$, and then to three maxima near $y=-14,0,14$. The electronic probability density and $|S_z^{(e)}|$ exhibit the same number of resolved transverse peaks, and the two ends retain opposite $z$-spin polarizations throughout this sequence.

\section{Conclusion}
\label{sec:conclusion}

In conclusion, we show that a finite-width altermagnetic strip can create a topological superconducting channel by Andreev confinement on a proximitized TI surface. The momentum-dependent exchange field shifts the particle- and hole-like branches near a non-$\bar{\Gamma}$ surface Dirac cone, enabling subgap propagation beneath the strip while the surrounding superconducting region provides transverse confinement. The resulting Andreev bands undergo successive transitions between trivial and topological superconducting phases, controlled at leading order by $J\cos(2\theta)$. This mechanism extends Andreev confinement to a compensated magnetic setting and links the topology of the confined channel to the strength and orientation of the $d$-wave spin-splitting pattern.

Within the lowest-pair description, we obtain asymptotic Majorana wave functions with opposite electronic spin polarizations at the two strip ends and a localization length determined by the ratio of the effective longitudinal velocity to the confined-band gap. The Majorana end states become increasingly delocalized as the confined-band gap closes near a topological transition. A $\pi/2$ rotation of the spin-splitting pattern reverses the spin polarization at each end. Beyond this description, numerical calculations reveal an evolution from single- to multiple-peaked transverse profiles across successive topological regions. Coherent mixing of transverse modes required by wave-function matching at the strip ends provides a physical interpretation of this spatial structure. Together, these results connect altermagnetic control of the topological spectrum with the localization and electronic spin texture of Majorana end states, and motivate spatially and spin-resolved tunneling studies of these heterostructures.

\begin{acknowledgments}
S.Z. was supported by the National Natural Science Foundation of China (Grant No. 12488101), HFNL Self-Deployed Project (Grant No. ZB2602000300), and Quantum Science and Technology-National Science and Technology Major Project (Grant No. 2021ZD0302801). \\
The authors used OpenAI's ChatGPT for assistance with brainstorming, code generation, and manuscript drafting. All AI-generated content was critically examined and independently verified by the authors.
\end{acknowledgments}

\appendix
\onecolumngrid

\section{Lattice-dispersion correction to the gap-closing condition}
\label{app:latticeboundary}

The main text uses the leading $k\cdot p$ theory near $\bar Y$ to obtain the gap-closing condition in Eq.~(\ref{eq:approx}). Here we improve the transverse propagation phase by retaining the trigonometric $k_y$ dependence inherited from the underlying lattice model at $k_x=0$, while retaining only the low-energy solutions associated with the \(\bar Y\) Dirac cone, rather than additional branches arising from the lattice regularization, and keeping the same $\Delta'=0$ limit. In this sense, the calculation incorporates higher-order lattice-dispersion corrections within the low-energy Andreev-quantization framework. At $k_x=0$, the normal-state dispersion underneath the strip in the eigensector of $s_1\tau_3$ with eigenvalue $\eta=\pm1$ and the eigensector of $\tau_3$ with eigenvalue $\lambda=\pm1$ is
\begin{equation}
    E_{\eta,\lambda}(k_y)=-\lambda\mu+\eta \lambda v\sin k_y+2\eta J_c(\cos k_y-1).
    \label{eq:appLatticeDispersion}
\end{equation}
For the two electron-like branches, $\lambda=+1$, the zero-energy equations can be written as
\begin{equation}
    v\sin k_y+2J_c\cos k_y=2J_c+\eta\mu,
    \label{eq:appLatticeElectronEq}
\end{equation}
where $J_c=J\cos(2\theta)$. Selecting the roots near $k_y=\pi$, Eq.~(\ref{eq:appLatticeElectronEq}) gives
\begin{equation}
    k_{e,\rightarrow/\leftarrow}=\pi-\arcsin\left(\frac{2J_c\mp\mu}{\sqrt{v^2+4J_c^2}}\right)-\arctan\left(\frac{2J_c}{v}\right).
    \label{eq:appLatticeRoots}
\end{equation}
Similarly, for the two hole-like branches, $\lambda=-1$, selecting the roots near $k_y=\pi$ gives
\begin{equation}
    k_{h,\leftarrow/\rightarrow}=\pi+\arcsin\left(\frac{2J_c\pm\mu}{\sqrt{v^2+4J_c^2}}\right)+\arctan\left(\frac{2J_c}{v}\right).
    \label{eq:appLatticeHoleRoots}
\end{equation}
These four roots determine the relative propagation phase accumulated by the electron- and hole-like modes across the strip. We introduce the momentum difference
\begin{equation}
    \Delta k= k_{h,\leftarrow}-k_{e,\rightarrow}=k_{h,\rightarrow}-k_{e,\leftarrow}=2\arctan\left(\frac{2J_c}{v}\right)+\arcsin\left(\frac{2J_c+\mu}{\sqrt{v^2+4J_c^2}}\right)+\arcsin\left(\frac{2J_c-\mu}{\sqrt{v^2+4J_c^2}}\right).
    \label{eq:appLatticeDeltaKMu}
\end{equation}
The two $\mu$-dependent terms enter symmetrically, so the correction linear in $\mu$ cancels. Neglecting the remaining quadratic correction in $\mu$, we set $\mu=0$. Using $\arcsin\left({2J_c}/{\sqrt{v^2+4J_c^2}}\right)=\arctan\left({2J_c}/{v}\right)$, we obtain
\begin{equation}
    \Delta k=4\arctan\left(\frac{2J_c}{v}\right).
\label{eq:appLatticeDeltaK}
\end{equation}

This momentum difference enters the round-trip propagation phase in the same way as the continuum momentum mismatch in Sec.~\ref{sec:andreev}. Using the same zero-energy Andreev reflection matrices, the bound-state condition becomes
\begin{equation}
    \Delta k\,W=(2n+1)\pi,\quad n\in\mathbb{Z}.
    \label{eq:appLatticeQuantization}
\end{equation}
Substituting Eq.~(\ref{eq:appLatticeDeltaK}) gives
\begin{equation}
    \frac{|J\cos(2\theta)|}{v}=\frac12\tan\left(\frac{(n+1/2)\pi}{2W}\right),\quad n\in\mathbb{N}.
    \label{eq:appLatticeBoundary}
\end{equation}
For $(n+1/2)\pi/W\ll1$, expanding the tangent gives
\begin{equation}
    \frac{|J\cos(2\theta)|}{v}=\frac{(n+1/2)\pi}{4W},\quad n\in\mathbb{N},
\end{equation}
which reproduces Eq.~(\ref{eq:approx}). Equation~(\ref{eq:appLatticeBoundary}) thus gives a lattice-dispersion-corrected phase boundary within the same low-energy Andreev-quantization framework as the main text.

\section{Finite superconducting $\Delta'$ correction to the gap-closing condition}
\label{app:finitepairing}

We retain the momentum-independent altermagnetic field $J_{\bar Y}\simeq-4J_c$ and allow a finite real pairing $0\leq\Delta'<4|J_c|$ underneath the strip, with the same superconducting phase as the exterior pairing $\Delta>0$. At $q_x=0$, the eigenvalue $\eta=\pm1$ of $s_1\tau_3$ is conserved, and the reduced Hamiltonian is
\begin{equation}
    H_\eta(q_y)=-4\eta J_c-(\eta vq_y+\mu)\rho_3+\Delta'\rho_1.
\end{equation}
We determine the interface reflection matrices by matching the propagating strip modes to the exterior decaying states, following the scattering construction of Ref.~\cite{PapajFu2021}.

The zero-energy dispersion equation is
\begin{equation}
    (\eta vq_y+\mu)^2+\Delta'^2=16J_c^2,
\end{equation}
giving
\begin{equation}
    q_{\eta,\pm}=-\frac{\eta}{v}\big(\mu\pm\sqrt{16J_c^2-\Delta'^2}\big).
    \label{eq:appFiniteDeltaMomenta}
\end{equation}
Thus propagating modes exist only for $4|J_c|>\Delta'$.

Labeling the roots by their electron- or hole-like limit as $\Delta'\to0$, their momenta and unnormalized spinors are
\begin{equation}
\begin{aligned}
    q_{e,\eta}&=-\frac{\eta\mu+\operatorname{sgn}(J_c)\sqrt{16J_c^2-\Delta'^2}}{v},
    &\chi_{e,\eta}&=\begin{pmatrix}
    1, \eta\tan\dfrac{\vartheta}{2}
    \end{pmatrix}^T,\\
    q_{h,\eta}&=-\frac{\eta\mu-\operatorname{sgn}(J_c)\sqrt{16J_c^2-\Delta'^2}}{v},
    &\chi_{h,\eta}&=\begin{pmatrix}
    \eta\tan\dfrac{\vartheta}{2},
    1
    \end{pmatrix}^T.
\end{aligned}
\label{eq:appFiniteDeltaSpinors}
\end{equation}
Here, we have defined for convenience the angle $\vartheta$ satisfying $\sin \vartheta= \Delta'/(4J_c)$ and $\cos \vartheta= \sqrt{16J_c^2-\Delta'^2}/|4J_c|$. Using the transverse current operator $-\eta v\rho_3$, we find that these spinors carry equal and opposite currents,
\begin{equation}
    j_{e,\eta}=-\dfrac{2\eta v\cos\vartheta}{1+\cos\vartheta},\quad j_{h,\eta}=-j_{e,\eta}.
\end{equation}
Dividing both spinors by the same $\sqrt{|j_{e,\eta}|}$ normalizes them to unit flux and leaves the matching amplitude ratios below unchanged. Thus $e_{\rightarrow}$ and $h_{\leftarrow}$ belong to $\eta=-1$, while $h_{\rightarrow}$ and $e_{\leftarrow}$ belong to $\eta=+1$. The relative momentum is
\begin{equation}
    \Delta q=q_{h,\leftarrow}-q_{e,\rightarrow}=q_{h,\rightarrow}-q_{e,\leftarrow}={2\operatorname{sgn}(J_c)}\sqrt{16J_c^2-\Delta'^2}/v.
    \label{eq:appFiniteDeltaMismatch}
\end{equation}
The chemical potential shifts both roots in a sector equally and therefore cancels from this difference.

Outside the strip, the zero-energy decaying solutions have momenta $q_y=-\eta\mu/v\pm i\Delta/v$ and boundary spinors
\begin{equation}
    \psi_\eta(-W/2)\propto\begin{pmatrix}1, -i\eta\end{pmatrix}^T,\quad \psi_\eta(W/2)\propto\begin{pmatrix}1,i\eta\end{pmatrix}^T.
    \label{eq:appFiniteDeltaBoundarySpinors}
\end{equation}
Near either interface $y=\pm W/2$, we may write the strip wave function as
\begin{equation}
    \psi_\eta(y)=a_e e^{iq_{e,\eta}(y\mp W/2)}\chi_{e,\eta}+a_h e^{iq_{h,\eta}(y\mp W/2)}\chi_{h,\eta}.
    \label{eq:appFiniteDeltaSuperposition}
\end{equation}
Continuity at $y=\pm W/2$ requires $a_e\chi_{e,\eta}+a_h\chi_{h,\eta}=c_\pm(1,\pm i\eta)^T$. Eliminating $c_\pm$ between the two components gives
\begin{equation}
    \left.\frac{a_h}{a_e}\right|_\pm=
    % \frac{\eta\left(\pm i\big(4|J_c|+\sqrt{16J_c^2-\Delta'^2}\big)-\operatorname{sgn}(J_c)\Delta'\right)}{4|J_c|+\sqrt{16J_c^2-\Delta'^2}\mp i\operatorname{sgn}(J_c)\Delta'}
    %\frac{\eta}{4|J_c|}\left(-\operatorname{sgn}(J_c)\Delta'\pm i\sqrt{16J_c^2-\Delta'^2}\right).
    \pm i \eta\; e^{\pm i \vartheta}
\label{eq:appFiniteDeltaMatching}
\end{equation}
The absence of a counterpropagating mode of the same type in a fixed $\eta$ sector forbids normal reflection. To use the channel convention of Eq.~(\ref{eq:Smatrix}), the upward and downward amplitude vectors are $(a_{e,-},a_{h,+})^T$ and $(a_{e,+},-a_{h,-})^T$, respectively. Equation~(\ref{eq:appFiniteDeltaMatching}) then gives
%\begin{equation}
%\begin{aligned}
%   S_+&=\frac{1}{4|J_c|}\begin{pmatrix}
%    0&-\operatorname{sgn}(J_c)\Delta'-i\sqrt{16J_c^2-\Delta'^2}\\
%    -\operatorname{sgn}(J_c)\Delta'+i\sqrt{16J_c^2-\Delta'^2}&0
%    \end{pmatrix},\\
%    S_-&=\frac{1}{4|J_c|}\begin{pmatrix}
%    0&-\operatorname{sgn}(J_c)\Delta'+i\sqrt{16J_c^2-\Delta'^2}\\
%    -\operatorname{sgn}(J_c)\Delta'-i\sqrt{16J_c^2-\Delta'^2}&0
%    \end{pmatrix}.
%\end{aligned}
%\label{eq:appFiniteDeltaReflection}
%\end{equation}
\begin{equation}
\begin{aligned}
    S_+&= \begin{pmatrix}
    0& -i e^{-i\vartheta}\\
    i e^{i\vartheta}&0
    \end{pmatrix},\qquad 
    S_- =\begin{pmatrix}
    0& i e^{i\vartheta}\\
    -i e^{-i\vartheta} &0
    \end{pmatrix}.
\end{aligned}
\label{eq:appFiniteDeltaReflection}
\end{equation}
Both matrices are unitary and reduce to Eq.~(\ref{eq:Smatrix}) at $\Delta'=0$.

One complete round trip is therefore given by
\begin{equation}
    S_-T_-S_+T_+=\operatorname{diag}\left((S_+)_{21}^{\,2}e^{-i\Delta qW},(S_+)_{12}^{\,2}e^{i\Delta qW}\right).
    \label{eq:appFiniteDeltaRoundTrip}
\end{equation}
Considering $4|J_c|>\Delta'$, substitution of the reflection amplitudes and Eq.~(\ref{eq:appFiniteDeltaMismatch}) gives
%\begin{equation}
%    \cos\left(\frac{W}{v}\sqrt{16J_c^2-\Delta'^2}\right)+\frac{\Delta'}{\sqrt{16J_c^2-\Delta'^2}}\sin\left(\frac{W}{v}\sqrt{16J_c^2-\Delta'^2}\right)=0,
%    \label{eq:appFiniteDeltaBoundary}
% \end{equation}
\begin{equation}
    \frac{W}{v}\sqrt{16J^2\cos^2(2\theta)-\Delta'^2}=\left(n+\frac12\right)\pi+\arcsin\left(\frac{\Delta'}{4|J\cos(2\theta)|}\right),\quad n\in\mathbb{N} .
    \label{eq:appFiniteDeltaBoundary2}
\end{equation}
For the limit $\Delta'=0$, this reduces to Eq.~(\ref{eq:approx}). A finite $\Delta'$ changes both the propagation momenta and the interface reflection amplitudes. The exterior gap $\Delta$ fixes the decay length $v/\Delta$, but does not enter the zero-energy reflection spinors or the resulting phase boundary.

\section{Single-Dirac-cone regularization correction to the gap-closing condition}
\label{app:singlecone}

The lattice Hamiltonian in Eq.~(\ref{eq:H0}) contains additional Dirac cones. Following the regularization used in Ref.~\cite{PapajFu2021}, we add
\begin{equation}
    H_W(\bm k)=rv(2-\cos k_x+\cos k_y)s_3\tau_3,
    \label{eq:WilsonY}
\end{equation}
where $r$ is dimensionless. The prefactor takes the values $2$, $4$, $0$, and $2$ at $\bar\Gamma=(0,0)$, $\bar X=(\pi,0)$, $\bar Y=(0,\pi)$, and $\bar M=(\pi,\pi)$, respectively. The unwanted cones thus acquire masses of order $rv$, while the target $\bar Y$ cone remains massless. Near $\bar Y$,
\begin{equation}
    H_W(\bm q)=\frac{rv}{2}(q_x^2+q_y^2)s_3\tau_3+O(q^4).
    \label{eq:WilsonYExpansion}
\end{equation}
We calculate its leading correction at $q_x=0$ and $\Delta'=0$, retaining $J_{\bar Y}\simeq-4J_c$. The interface calculation below uses this quadratic continuum Hamiltonian, with the Wilson coefficient uniform in the full system.

Inside the strip,
\begin{equation}
    H_{\bar Y}(0,q_y)=-\mu\tau_3-vq_y s_1-4J_c s_1\tau_3+\frac{rv}{2}q_y^2s_3\tau_3.
    \label{eq:WilsonStripHamiltonian}
\end{equation}
The electron and hole blocks remain separate. Their zero-energy momenta obey $(vq_y\pm4J_c)^2+{r^2v^2}q_y^4/4=\mu^2$,
respectively. For $\mu>0$ and $|r|(\mu+4|J_c|)^2\ll\mu v$, expanding about the four real roots at $r=0$ gives
\begin{equation}
    q_{e,\rightarrow/\leftarrow}\simeq\frac{-4J_c\pm\mu}{v}\mp\frac{r^2(-4J_c\pm\mu)^4}{8\mu v^3},\quad q_{h,\leftarrow/\rightarrow}\simeq\frac{4J_c\pm\mu}{v}\mp\frac{r^2(4J_c\pm\mu)^4}{8\mu v^3}.
\label{eq:WilsonPerturbedMomenta}
\end{equation}
Subtracting the counterpropagating electron and hole momenta gives
\begin{equation}
    \Delta q=q_{h,\leftarrow}-q_{e,\rightarrow}=q_{h,\rightarrow}-q_{e,\leftarrow}\simeq\frac{8J_c}{v}\left[1-\frac{r^2}{2v^2}(16J_c^2+\mu^2)\right].
    \label{eq:WilsonMomentumMismatch}
\end{equation}

The Wilson term also rotates the spinors, so we next calculate the interface reflection. In the $s_1$ eigenbasis $|+\rangle=(1,1)^T/\sqrt2$, $|-\rangle=(1,-1)^T/\sqrt2$, the electron and hole spinors obey the zero-mode equations in each $\tau_3$ sector as
\begin{equation}
    \begin{pmatrix}
    -\mu-vq-4J_c & rvq^2/2\\
    rvq^2/2 & -\mu+vq+4J_c
    \end{pmatrix}u_e(q)=0,\quad\begin{pmatrix}
    \mu-vq+4J_c & -rvq^2/2\\
    -rvq^2/2 & \mu+vq-4J_c
    \end{pmatrix}u_h(q)=0.
\label{eq:WilsonStripSpinors}
\end{equation}
The four-component modes are $U_e=(u_e,0)^T$ and $U_h=(0,u_h)^T$, in the order $(e+,e-,h+,h-)$. The electron and hole equations give
\begin{equation}
    u_{e,\rightarrow}=\begin{pmatrix}
    \dfrac{rvq_{e,\rightarrow}^2}{2(\mu+vq_{e,\rightarrow}+4J_c)}, 
    1
    \end{pmatrix}^T,\quad u_{h,\leftarrow}=\begin{pmatrix}
    1, 
    \dfrac{rvq_{h,\leftarrow}^2}{2(\mu+vq_{h,\leftarrow}-4J_c)}
    \end{pmatrix}^T.
\end{equation}
The leading correction yields
\begin{equation}
    u_{e,\rightarrow}\simeq\begin{pmatrix}
    \dfrac{r(\mu-4J_c)^2}{4\mu v}, 1
    \end{pmatrix}^T,\quad u_{h,\leftarrow}\simeq\begin{pmatrix}
    1, \dfrac{r(\mu+4J_c)^2}{4\mu v}
    \end{pmatrix}^T.
    \label{eq:WilsonSlowSpinors}
\end{equation}

Place the upper interface at $y=0$, with the strip at $y<0$. The exterior solutions follow from the BdG equation with $J_c=0$ and pairing $\Delta$, retaining only modes that decay into the superconductor. For an incoming electron, the strip wave function contains the incident electron, the reflected electron and hole, and the additional modes introduced by the quadratic Wilson term that decay into the strip.

Since the Wilson coefficient is uniform across the interface, both the wave function and its first derivative are continuous, $\psi(0^-)=\psi(0^+)$ and $\partial_y\psi(0^-)=\partial_y\psi(0^+)$.
These conditions give linear equations for the reflection amplitudes and the coefficients of the decaying modes. Solving these equations and expanding the reflection amplitudes through second order in the low-energy expansion gives
\begin{equation}
    \widetilde r_{ee}=0,\quad \widetilde r_{he}\simeq i\left[1-\frac{r^2J_c}{v^2}\left(5\mu+\frac{16J_c^2}{\mu}\right)\right]-\frac{2r^2J_c\Delta}{v^2}.
    \label{eq:WilsonUnnormalizedReflection}
\end{equation}
The tilde indicates that the strip spinors have not yet been normalized to unit flux. Using
$j_y=U^\dagger(-v s_1+rvq_y s_3\tau_3)U$ with Eq.~(\ref{eq:WilsonSlowSpinors}), the normalized reflection amplitudes are therefore simplified to
\begin{equation}
    r_{he}=\widetilde r_{he}\sqrt{\frac{|j_h|}{|j_e|}}\simeq\widetilde r_{he}\left(1+\frac{r^2J_c}{v^2}\left(5\mu+\frac{16J_c^2}{\mu}\right)\right)\simeq i\left(1+\frac{2ir^2J_c\Delta}{v^2}\right)\simeq i\exp\left(\frac{2ir^2J_c\Delta}{v^2}\right).
    \label{eq:WilsonAndreevPhase}
\end{equation}
Similarly, the hole-incidence calculation gives $r_{eh}=r_{he}^*$, while reversing the decaying directions gives the lower-interface amplitudes. In the channel convention of Eq.~(\ref{eq:Smatrix}),
\begin{equation}
    S_+\simeq\begin{pmatrix}
        0&-i e^{-2ir^2J_c\Delta/v^2}\\
        i e^{2ir^2J_c\Delta/v^2}&0
    \end{pmatrix},
    \qquad
    S_-\simeq\begin{pmatrix}
        0&i e^{2ir^2J_c\Delta/v^2}\\
        -i e^{-2ir^2J_c\Delta/v^2}&0
    \end{pmatrix}.
    \label{eq:WilsonReflectionMatrices}
\end{equation}

For $W\gg|r|$, the additional decaying modes inside the strip have negligible overlap between opposite interfaces. The propagation matrices therefore involve only the four propagating strip modes in Eq.~(\ref{eq:WilsonPerturbedMomenta}). Combining them with Eq.~(\ref{eq:WilsonReflectionMatrices}) gives
\begin{equation}
    S_-T_-S_+T_+\simeq-\operatorname{diag}\left(e^{-i(\Delta qW-4r^2J_c\Delta/v^2)},e^{i(\Delta qW-4r^2J_c\Delta/v^2)}\right).
\end{equation}
The bound-state condition therefore reduces to
\begin{equation}
    \det(\mathbb I-S_-T_-S_+T_+)\simeq4\cos^2\left(\frac{\Delta qW}{2}-\frac{2r^2J_c\Delta}{v^2}\right)=0,
    \label{eq:WilsonQuantization}
\end{equation}
which includes both the propagation correction and the interface phase derived above. The gap-closing condition can therefore be written as
\begin{equation}
    \frac{4|J_c|W}{v}-\frac{2r^2|J_c|W}{v^3}(16J_c^2+\mu^2)-\frac{2r^2|J_c|\Delta}{v^2}\simeq\left(n+\frac12\right)\pi,\quad n\in\mathbb{N}.
    \label{eq:WilsonTotalPhase}
\end{equation}
Solving this condition perturbatively gives the phase boundary
\begin{equation}
    \frac{|J\cos(2\theta)|}{v}\simeq\frac{(n+1/2)\pi}{4W}\left(1+\frac{r^2}{2}\left(\frac{\mu^2}{v^2}+\frac{\pi^2(n+1/2)^2}{W^2}+\frac{\Delta}{vW}\right)\right).
    \label{eq:WilsonBoundaryTheta}
\end{equation}
Setting $r=0$ reproduces Eq.~(\ref{eq:approx}). Around its $n$-th boundary
\begin{equation}
    \frac{|J_c|}{v}\simeq\frac{(n+1/2)\pi}{4W}
\end{equation}
and within the approximation $ E_*=\max\left\{|\mu|,4J,\Delta,\Delta'\right\}\ll v$, the relative displacement is given by
\begin{equation}
    \frac{\delta J_c}{J_c}\simeq\frac{r^2}{2}\left(\left(\frac{\mu}{v}\right)^2+\left(\frac{4J_c}{v}\right)^2+\frac{1}{(n+1/2)\pi}\frac{\Delta}{v}\frac{4|J_c|}{v}\right)\simeq O\left(\left(\frac{rE_*}{v}\right)^2\right).
\end{equation}
Wilson regularization therefore shifts the gap-closing boundaries
toward larger $|J\cos(2\theta)|$ by a second-order relative amount.

\section{Finite-$\Delta'$ correction to the transverse electronic density of a confined state}
\label{app:finitepairingprofile}

The preceding Appendix~\ref{app:finitepairing} considered the effect of a finite pairing $\Delta'$ underneath the strip on the zero-energy gap-closing condition. Here we use the same constant-altermagnet approximation, $J_{\bar Y}\simeq-4J_c$, to determine the distinct correction that $\Delta'$ produces in the transverse electronic probability density of a confined state at a finite subgap energy $|\epsilon|<\Delta$. This correction is useful for separating the oscillatory structure caused by electron-hole mixing inside the strip from the multi-peaked end-state profiles discussed in Sec.~\ref{sec:doublepeak}.

At $q_x=0$, the reduced Hamiltonian in the eigensector of $s_1\tau_3$ with eigenvalue $\eta=\pm1$ remains
\begin{equation}
    H_\eta(q_y)=-4\eta J_c-(\eta vq_y+\mu)\rho_3+\Delta'\rho_1.
    \label{eq:appProfileHamiltonian}
\end{equation}
For a state at energy $\epsilon$, the BdG equation gives
\begin{equation}
    (\eta vq_y+\mu)^2+\Delta'^2=(\epsilon+4\eta J_c)^2.
    \label{eq:appProfileDispersion}
\end{equation}
Hence propagating solutions inside the strip exist when $|\epsilon+4\eta J_c|>\Delta'$. Their two transverse momenta are
\begin{equation}
    q_{\eta,\pm}(\epsilon)=-\frac{\eta}{v}\left[\mu\pm\sqrt{(\epsilon+4\eta J_c)^2-\Delta'^2}\right].
    \label{eq:appProfileMomenta}
\end{equation}
At $\epsilon=0$, Eq.~(\ref{eq:appProfileMomenta}) reduces to Eq.~(\ref{eq:appFiniteDeltaMomenta}).

For definiteness, consider a branch with $\epsilon+4\eta J_c>\Delta'$. The two normalized BdG spinors can be chosen as
\begin{subequations}
\begin{align}
    \chi_{\eta,+}=\frac{1}{\sqrt{2(\epsilon+4\eta J_c)\left(\epsilon+4\eta J_c+Q\right)}}\begin{pmatrix}
    \epsilon+4\eta J_c+Q\\
    \Delta'
    \end{pmatrix},
    \label{eq:appProfileSpinorPlus} \\
    \chi_{\eta,-}=\frac{1}{\sqrt{2(\epsilon+4\eta J_c)\left(\epsilon+4\eta J_c+Q\right)}}\begin{pmatrix}
    \Delta'\\
    \epsilon+4\eta J_c+Q
    \end{pmatrix},
    \label{eq:appProfileSpinorMinus}
\end{align}
\end{subequations}
where $Q=\sqrt{(\epsilon+4\eta J_c)^2-\Delta'^2}$. The branch with $\epsilon+4\eta J_c<-\Delta'$ is obtained analogously and gives the same magnitude of the density modulation below. For the electronic probability density, the relevant overlap includes the projector $P_e=(1+\rho_3)/2$ in the reduced sector. Directly from Eqs.~(\ref{eq:appProfileSpinorPlus}) and (\ref{eq:appProfileSpinorMinus}), we find
\begin{equation}
    \chi_{\eta,+}^{\dagger}P_e\chi_{\eta,-}
    =\frac{\Delta'}{2(\epsilon+4\eta J_c)}.
    \label{eq:appProfileSpinorOverlap}
\end{equation}
Thus the electron-hole mixing produced by $\Delta'$ allows the two propagating components to interfere in the electronic probability density.

Inside the strip, a confined state in this $\eta$ sector is a superposition
\begin{equation}
    \psi_\eta(y)=A e^{iq_{\eta,+}y}\chi_{\eta,+}+B e^{iq_{\eta,-}y}\chi_{\eta,-}.
    \label{eq:appProfileWavefunction}
\end{equation}
For a transversely confined state, the quasiparticle current vanishes because the wave function decays outside the strip. Using the normalized spinors above gives
\begin{equation}
    j_y=-\eta v\psi_\eta^\dagger\rho_3\psi_\eta
    =-\frac{\eta vQ}{\epsilon+4\eta J_c}
    \left(|A|^2-|B|^2\right)=0,
\end{equation}
so $|A|=|B|$. Using Eq.~(\ref{eq:appProfileSpinorOverlap}), its electronic density is therefore
\begin{equation}
    n=\psi_\eta^{\dagger}(y)P_e\psi_\eta(y)
    =\frac{|A|^2+|B|^2}{2}
    +\frac{\Delta'}{\epsilon+4\eta J_c}
    \mathrm{Re}\big[A^*B e^{-i(q_{\eta,+}-q_{\eta,-})y}\big].
    \label{eq:appProfileDensity}
\end{equation}
Equation~(\ref{eq:appProfileDensity}) shows that finite $\Delta'$ produces a transverse cosine with wave vector
\begin{equation}
    K_{\Delta'}=|q_{\eta,+}-q_{\eta,-}|
    = 2\sqrt{(\epsilon+4\eta J_c)^2-\Delta'^2}/v.
    \label{eq:appProfileWavevector}
\end{equation}
Writing $n_0=(|A|^2+|B|^2)/2$ for the nonoscillating electronic density, the relative modulation satisfies
\begin{equation}
    \left|\frac{\delta n_{\Delta'}}{n_0}\right|
    \leq\left|\frac{\Delta'}{\epsilon+4\eta J_c}\right|.
    \label{eq:appProfileAmplitude}
\end{equation}

In the regime relevant to the higher-$J$ topological regions, $4|J_c|\gg|\epsilon|,\Delta'$, Eqs.~(\ref{eq:appProfileWavevector}) and (\ref{eq:appProfileAmplitude}) reduce to
\begin{equation}
    K_{\Delta'}\simeq\frac{8|J_c|}{v},\quad
    \left|\frac{\delta n_{\Delta'}}{n_0}\right|
    \lesssim\frac{\Delta'}{4|J_c|}.
    \label{eq:appProfileLargeJ}
\end{equation}
Therefore, increasing $|J_c|$ makes the finite-$\Delta'$ modulation shorter in wavelength while reducing its relative amplitude. Since the multi-peaked end-state profiles discussed in Sec.~\ref{sec:doublepeak} are obtained at $\Delta'=0$, this additional finite-$\Delta'$ modulation does not account for their formation.

\end{document}